\documentclass[useAMS,usenatbib]{mn2e}
\usepackage{times}
\usepackage{graphicx}

\bibpunct{(}{)}{;}{a}{}{,}

\def\beq{\begin{equation}}
\def\eeq{\end{equation}}
\def\like{\ensuremath{\mathcal{L}}}
\def\deg{\ensuremath{^{\circ}}}

\newcommand{\figce}[5]{ 
  \begin{figure*}
    \includegraphics[angle=0,width=\columnwidth]{#1}
    \hspace*{5pt}
    \includegraphics[angle=0,width=\columnwidth]{#2}
    \caption{ {\sl (Left) } #3 {\sl (Right) } #4 \label{#5}}
  \end{figure*}
  }

\title[Robust statistical measures of single stellar populations]{
  A robust statistical estimation of the basic parameters of single 
  stellar populations.  I. Method}

\author[X. Hernandez and D. Valls--Gabaud]
{X. Hernandez$^{1}$ and David Valls--Gabaud$^{2}$ \\
$^1$ Instituto de Astronom\'{\i}a, Universidad Nacional Aut\'onoma de M\'exico
     A. P. 70--264,  M\'exico 04510 D.F., M\'exico\\
$^2$ GEPI, CNRS UMR 8111, Observatoire de Paris,
     5, Place Jules Janssen, 92195 Meudon Cedex, France
}

\begin{document}

\date{Accepted ... Received ... ; in original form ... }

\pagerange{\pageref{firstpage}--\pageref{lastpage}} \pubyear{2007}

\maketitle

\label{firstpage}

\begin{abstract}
The colour-magnitude diagrams of resolved single stellar populations, such as open and 
globular clusters, have provided the best natural laboratories to test stellar evolution theory. 
Whilst a variety of techniques have been used to infer the basic properties of these simple 
populations, systematic uncertainties arise from the purely geometrical degeneracy produced by the 
similar shape of isochrones of different ages and metallicities. Here we present an objective 
and robust statistical technique which lifts this degeneracy to a great extent through the 
use of a key observable: the number of stars along the isochrone. Through extensive Monte 
Carlo simulations we show that, for instance, we can infer the four main parameters (age, 
metallicity, distance and reddening) in an objective way, along with robust confidence intervals
and their full covariance matrix. We show that systematic uncertainties due to field contamination,
unresolved binaries, initial or present-day stellar mass function are either negligible or well under 
control. This technique provides, for the first time, a proper way to infer with unprecedented accuracy 
the fundamental properties of simple stellar populations, in an easy-to-implement algorithm. 
\end{abstract}


\section{Introduction}

The theory of stellar evolution has been tested extensively mainly through well-studied
binary stars ({\sl e.g.}, Popper 1980, Lastennet \& Valls-Gabaud 2002, Ribas 2006) 
or simple, coeval stellar populations such as open and
globular clusters ({\sl e.g.} Vandenberg et al. 1996, Renzini \& Fusi Pecci 1988). 
It has been extraordinarily successful at explaining most 
stellar properties across the different evolutionary stages, by incorporating increasingly
complex physics in both stellar interiors and atmospheres (see Salaris \& Cassisi 2006 
for an updated and extensive review). In contrast, and perhaps somewhat
paradoxically, the determination of the fundamental parameters 
of single stellar populations, such as distance, age,
metallicity, reddening, convective parameters, etc, has seldom been the subject of
a rigorous mathematical determination. More often than not, isochrones are ``fitted''
by eye, and very rough estimates of errors are given, if given at all. The reason
for this state of affairs is perhaps the many degeneracies between these
parameters, which makes it seemingly impossible to find a unique solution to the
set of fundamental parameters.  Alternative, semi-empirical methods have been developed instead
to infer {\sl relative} quantities, rather than absolute ones. For instance, the
so-called horizontal method relies on the difference in colour between the turn-off (TO) 
point and the base of the red giant branch (RGB), and is independent of the assumptions
made on the convective theory. The vertical method, on the other hand, relies on the
difference in apparent magnitude between horizontal branch (HB) stars and turn-off stars. Both
methods rely heavily on ``semi-empirical calibrations'' provided by stellar models, and are subject
to many uncertainties ({\sl e.g.}, How to define properly the TO point when photometric errors
blur this region of the colour-magnitude diagram ? What is the vertical offset HB--TO
when the HB is not horizontal ?) which nevertheless have been tackled with some
success ({\sl e.g.} Rosenberg et al. 1999, Salaris \& Weiss 2002).

Here we want to point out that many of the degeneracies are actually mere artifacts, and that
a proper and rigorous mathematical technique can be formulated to measure these parameters
in a robust and objective way. For example, the well-known age-metallicity degeneracy,
which states that a given isochrone can be fitted with a different age provided the
metallicity is changed appropriately, is a purely {\sl geometrical} degeneracy : the {\sl shape}
of the isochrones is the same, yet stellar evolution predicts that stars with a
different metal content will evolve at a different {\sl speed}. Hence, the age-metallicity
degeneracy is not a physical one, but simply a result of using the geometrical shape
of isochrones as the only discriminant factor. Clearly, a simple statistical estimator
that would count the number of stars {\sl along} the isochrones would give very
different results for any pair of age/metallicity values that would give the same geometrical
shape to the isochrones. In a more formal way, let us consider a curvilinear coordinate $s$ along an 
isochrone of given parameters (say, distance, reddening, metallicity, age, alpha elements enhancement, etc). 
This position depends only on the age $t$ of that isochrone
and on the mass $m$ of the star at that precise locus, so that $s=s(m,t)$ only. An offset in
age $dt$ is reflected only through changes in mass $m$ and position $s$ along the isochrone of
age $t$ since 
\beq
dt(m,s) \; = \; \left.\frac{ \partial t }{\partial m}\right|_s \, dm \; + \;
             \left.\frac{ \partial t }{\partial s}\right|_m \, ds \qquad .
\eeq
For a star in this isochrone the offset is, by definition, zero, hence 
\beq
\left.\frac{ \partial m }{\partial s}\right|_t \; = \; - \; 
\left.\frac{ \partial m }{\partial t}\right|_s \; \times \;
\left( \left.\frac{ \partial s }{\partial t}\right|_m \right)^{-1}   \quad .
\label{eq2}
\eeq
The first term is always finite.
The second term is the evolutionary speed, the rate of change for a given mass $m$ 
of its coordinate along the isochrone when the age changes by some small amount. One can
think of $s$ as representing some evolutionary phase, and so this term will be large
when the phase is short-lived : a small variation in age yields a very large
change in position along the isochrone. Alternatively, for a given age, and since
the first term is always finite, a wide variation in position implies a narrow
range in mass. This is the case of the red giant branch or the white dwarf
cooling sequence\footnote{excluding the bottom of the cooling sequence, where white dwarfs
having a large range of progenitor masses pile up}, for instance. On the other hand, slowly evolving phases such
as the main sequence have small evolutionary speeds and wide ranges in mass
for a given interval along an isochrone. Clearly, the most important phases to
discriminate between alternative ages and metallicities will be the post main-sequence
ones, where the range of mass is small (and hence insensitive to the details of the
stellar mass function), and at the same time where evolutionary speeds are large. 
If we consider the mid- to lower main sequence, at fixed metallicity, isochrones of all ages trace the same locus,
with only marginal changes in the density distribution of points amongst them. In this sense, one
of the parameters, $t$, is to a large extent absent from the main sequence,
while phases beyond are always substantially a function of both age and metallicity.
The density of stars along an isochrone is therefore
\beq
\frac{d N}{d s} \; = \; - \left(\frac{dN}{dm}\right) \times 
\left( \left. \frac{\partial m}{\partial t}\right|_s \right) \times
\left( \left. \frac{\partial t}{\partial s}\right|_m \right) \qquad .
\eeq
The first term is related to  the initial stellar mass function (IMF), and, as discussed above,
the second term is finite while the third is a strong function of the evolutionary
speed. If the mass after the turn-off is assumed to be roughly constant, this
implies that the {\sl ratio} in the number of stars in two different evolutionary
stages after the turn-off will only depend on the ratio of their evolutionary
time scales\footnote{In the context of stellar population synthesis, this is
known as the fuel consumption theorem (Renzini \& Buzzoni 1983).}.

In terms of observables, this well-know relation has been exploited using luminosity
functions as a proxy (as pioneered by Paczy\'nski 1984), that is, number counts of stars as a function of
apparent magnitude. This is not the isochrone coordinate $s$, but rather the
projection of the isochrone on the vertical axis of the colour-magnitude diagram. 
The fact that the sub-giant branch often appears nearly horizontal in CMDs implies
that luminosity bins in this regime will not be discriminant. This is unfortunate because this
phase is more heavily populated than the red giant branch, and thus the use of
luminosity functions is not optimal. Likewise, the projection on the horizontal
axis, a colour, is less sensitive to the details along the RGB. Clearly, the
optimal strategy is to use the full information provided by the observed density
of stars along the isochrone. 

To summarize existing methods, one approach has grown out of the fitting of
'geometrical' indicators of the observed CMD to isochrones, from single particular
indicators such as turn off point (TO), magnitude or colour difference between two 
points on the CMD, TO vs. zero-age horizontal branch or tip of the RGB and RGB bump
position and combinations thereof. Examples can be found in Renzini \& Fusi Pecci (1988),
Buonanno et al. (1998), Salaris \& Cassisi (1998), Salaris \& Weiss (1998), VandenBerg \& Durrel (1990),
Ferraro et al. (1999) and Rosenberg et al. (1999).
The accuracy of these approaches has been extended by including 
several such geometrical indicators simultaneously (e.g. VandenBerg 2000, 
Meissner \& Weiss 2006), or a complete geometrical comparison between observed CMD
and isochrone (e.g. Straniero \& Chieffi 1991). The reduction of the CMD to a few numbers,
given the varied and complex physics which enters in determining them implies using only
a subset of the information available in the CMD, while the unavoidable observational errors
present necessarily lead to ambiguities in the identification of the key points to be used, which
are hard to quantify formally, and which result in uncertain confidence intervals being assigned to
the inferences derived. Neglecting stellar evolutionary effects limits the information content of 
the isochrones against which the CMDs are compared.

On the other hand, approaches concentrating explicitly on the stellar evolutionary effects  
have grown out of the already mentioned analysis of luminosity functions presented by
Paczy\'nski (1984), a projection of the CMD onto the vertical axis. 
Extensions have mostly centered on the easy to measure and 
age sensitive features of the RGB luminosity function ({\sl e.g.} Jimenez \& Padoan, 
1998, and Zoccali \& Piotto 2000).
Here again, observational errors together with the binning adopted in generating luminosity 
functions, imply a limited use of the information available in the observation, while
not including the information encoded by the geometry of both the CMD and isochrones 
limits the accuracy of the inferences.

There are have been a few attempts in the past at using the full information available in a
colour-magnitude diagram, as pioneered by Flannery \& Johnson (1982) and further
developed by Wilson \& Hurrey (2003) and Naylor \& Jeffries (2006). Unfortunately all
these approaches, while mathematically correct, fail to include properly the key
discriminating factor discussed above, the stellar evolutionary speed, resulting in
wide uncertainties in the inferred parameters. The only other attempt we are aware
of, in the context of inferring parameters from a stellar population, was made by
J{\o}rgensen \& Lindegren (2005), who adapted our formalism developed in a previous
paper (Hernandez, Valls-Gabaud \& Gilmore 1999) to derive stellar ages. The present
paper is a further and robust extension of this approach. Paper II in this series
(Valls-Gabaud \& Hernandez 2007) will apply the method to different sets of
observations of stellar populations.
         
In section 2 we present the rigorous probabilistic model, 
which is tested with synthetic cases in Section 3. In section 4 we explore and quantify 
the effects of systematic uncertainties, in section 5 we present an application of the method to 
a real case, NGC~3201.  Finally in section 6 we present our 
conclusions on this novel approach.

\section{Probabilistic model}

In trying to recover the physical parameters of an observed cluster, we begin by constructing 
a model where only 4 numbers determine the above, the age, metallicity, distance and reddening 
of the cluster, a vector $(T, Z, D, R)$. Henceforth, we will define $D$ as the total
vertical displacement of the stars in the observed CMD with respect to their original
positions. This vertical displacement is of course the distance modulus and the contribution
of the extinction in the observed band, that is, $D = \mu + R_V \times \, R$, where $R$ is
the colour excess in the bands considered and $R_V$ the ratio of selective to total
absorption. Note that these contributions are, in general, correlated, since at larger
distances the reddening usually increases within the Galaxy. Note also that $R_V$ may depend
on the line of sight considered, even though a universal extinction curve is usually assumed.
For these reasons it is  simpler (and model-independent) 
to consider a vertical offset, and a horizontal
one, the colour excess, as two independent quantities. 

Thus, we assume the IMF to be known, the binary fraction to be
zero and the contamination of background/foreground stars  to be negligible.
This conditions are of course not true in a real case, and we will therefore treat them as
systematics, the effects of which we shall explore in detail in section 4.

From a Bayesian perspective, given an actual observed CMD, we want to identify the  
vector $(T, Z, D, R)$ which has the highest probability in having resulted 
in the actual observed cluster, as that from which the CMD most probably actually 
came from. This model is essentially what was introduced and tested in Hernandez et al. (1999),
and subsequently used in Hernandez et al. (2000a, 2000b) to infer star formation 
histories of resolved galaxies, although much simplified here for the treatment of single stellar populations. 
We must construct a statistical method for determining what is the probability
 that an observed CMD might have been the result of a given vector $(T, Z, D, R)$.
An observed CMD will be defined by a set of $i=1\, \cdots\, N_{\star}$ stars, each having a measured magnitude and 
colour (in whichever photometric bands or filters)  $(L_{i}, C_{i})$, with $\sigma(L_i)$ and  $\sigma(C_i)$ being
the measured (deduced from photometry data reduction procedures) 1-sigma errors in these
quantities. Errors in magnitudes and colours will be assumed as being Gaussian in nature,
and uncorrelated to each other for the time being. Hence, a reported star $(L_{i}, C_{i}; \sigma(L_i)$, $\sigma(C_i))$,
will be thought of as a two-dimensional Gaussian probability density distribution, 
centered on $(L_{i}, C_{i})$, with dispersions $\sigma(L_i)$, and $\sigma(C_i)$.

We begin with the probability that 
a given observed star, of magnitude and colour $(L_{i},C_{i})$, is in fact a real star
of some age $t$, and some metallicity, $Z$, of mass $m$, luminosity $L_{m}$ 
and colour $C_{m}$, observed with an assumed
vertical offset of $D$, and a reddening of $R$, both in units of magnitudes. 
This will be given by

\begin{eqnarray}
\lefteqn{G_{i}(T,Z,D,R,m) \; = }\nonumber \\
& & \left( {{1} \over {\displaystyle \sigma_{L}(m,i) \sigma_{C}(m,i)}}  \right) \, 
\times \nonumber \\
& &\times \, \exp\left( {{\displaystyle -D^{2}_{L}(m,i)} \over {\displaystyle 2 \sigma_{L}^2(m,i)}} \right) \times 
\exp\left( {{\displaystyle -D^{2}_{C}(m,i)} \over {\displaystyle 2 \sigma_{C}^2(m,i)}} \right) \quad . 
\label{eq:G}
\end{eqnarray}

In the above expression $D_{L}(m,i)$ and $D_{C}(m,i)$ are the differences 
in magnitude and
colour between the $i$th observed star and a generic star of age, metallicity, distance
modulus, reddening and mass $(T,Z,D,R;m)$. Also, 
\beq
\sigma_{L}^{2}(m,i)=\sigma(L_i)^{2}+\sigma(L_m)^{2} \; , \; 
\sigma_{C}^{2}(m,i)=\sigma(C_i)^{2}+\sigma(C_m)^{2}  ,
\eeq
where $\sigma(L_m)^{2}$ and $\sigma(C_m)^{2}$ are the errors which have to be applied to 
the theoretical comparison star of mass $m$ to map in onto the observed CMD. i.e., 
each theoretical star is being thought of as a Gaussian probability density distribution
centered on $(L_{m}, C_{m})$, with dispersions $\sigma(L_m)$ and $\sigma(C_m)$. 
Equation \ref{eq:G} 
is hence equivalent to the convolution over the entire CMD plane of two Gaussian probability 
distributions, a first one coming from the
reported observed star, and a second one from the probabilistic transfer function which we must apply to
a theoretical model star coming from an isochrone to map it onto a particular observed CMD.

If we wanted to consider model stars as mathematical points in this comparison, following
Bayes theorem, we would identify the probability that a particular reported real star, itself a 2-D probability
distribution, was in fact a certain model star, with the probability that this particular 
model star might actually come from the 2-D probability distribution of the reported star.
To treat the problem rigorously, we consider only the comparison between a 
reported star and a model one, only after this last one has been mapped onto the plane where the 
observed star exists, through a probability function, as described above.

In Equation \ref{eq:G} a numerical normalization constant has been left out, as in any case
 likelihood
functions have no absolute normalizations and only relative values are meaningful.
Whereas  $\sigma(L_i)$ and $\sigma(C_i)$ are supplied by a particular observation of a real
CMD, $\sigma(L_m)$ and $\sigma(C_m)$ have to be estimated. In a practical
application, one could obtain the average of  $\sigma(L_i)$ and $\sigma(C_i)$, as functions
of magnitude, and use those two functions as models for  $\sigma(L_m)$ and $\sigma(C_m)$.

Having the probability that a single observed star is in fact a particular model star, we can now
compute the probability that our observed star came from any point along a particular isochrone 
in our parameter space $(T, Z, D, R)$ as:

\begin{equation}
G_{i}(T,Z,D,R)  = 
\int\limits_{m_{0}(T,Z)}^{m_{1}(T,Z)}\rho(m;T,Z) G_{i}(T,Z,D,R,m) \, dm \, .
\label{eq:Gint}
\end{equation}

In the above expression $\rho(m;T,Z)$ is the density of stars along the isochrone
of metallicity $Z$ and age $T$, around the test star of mass $m$. As explained in \S1,
this density depends both on the assumed IMF and on the evolutionary speed predicted by
the stellar evolution code at that position of the CMD.  
The density of points along the isochrone, within a certain interval around a certain 
mass, in the highly
discriminating region beyond the main sequence, will be determined essentially by
the time spent by a star within that interval, as given
by the stellar evolutionary codes. If a large portion of the main sequence is included in the
study, the IMF will dominate $\rho(m;T,Z)$ over that region, as discussed in \S1. On the other hand, the mass interval
beyond the turn off point is so small, that practically any conceivable IMF will yield
identical results for $\rho(m;T,Z)$ in this region. The RBG has a lower density of points than
the main sequence not because of the IMF, but because of the fast stellar 
evolutionary speeds. 

In practice, the extremely rapid variations in magnitude and colour with mass
over critical regions of the isochrones make it convenient to carry out the 
integration in Equation \ref{eq:Gint} 
over a parameter measuring length over the isochrone, rather than mass, in which case $\rho(m;T,Z)$
must include weighting factors due to the relative duration of different stellar phases.
Also, $m_0(T,Z)$ and $m_1(T,Z)$ are a minimum and a maximum mass considered along each isochrone, such that 
the luminosity completeness limit of the observations always corresponds to the luminosity of
star $m_0(T,Z)$, while $m_1(T,Z)$ corresponds to the star of age and metallicity $(T,Z)$
at the tip of the RGB. The explicit exclusion of phases beyond the helium flash reduces to a
certain extent the resolution of the method, but increases significantly its robustness, as
it guarantees (see section 4) that only well-established physics are considered in the stellar evolutionary
models.

\figce{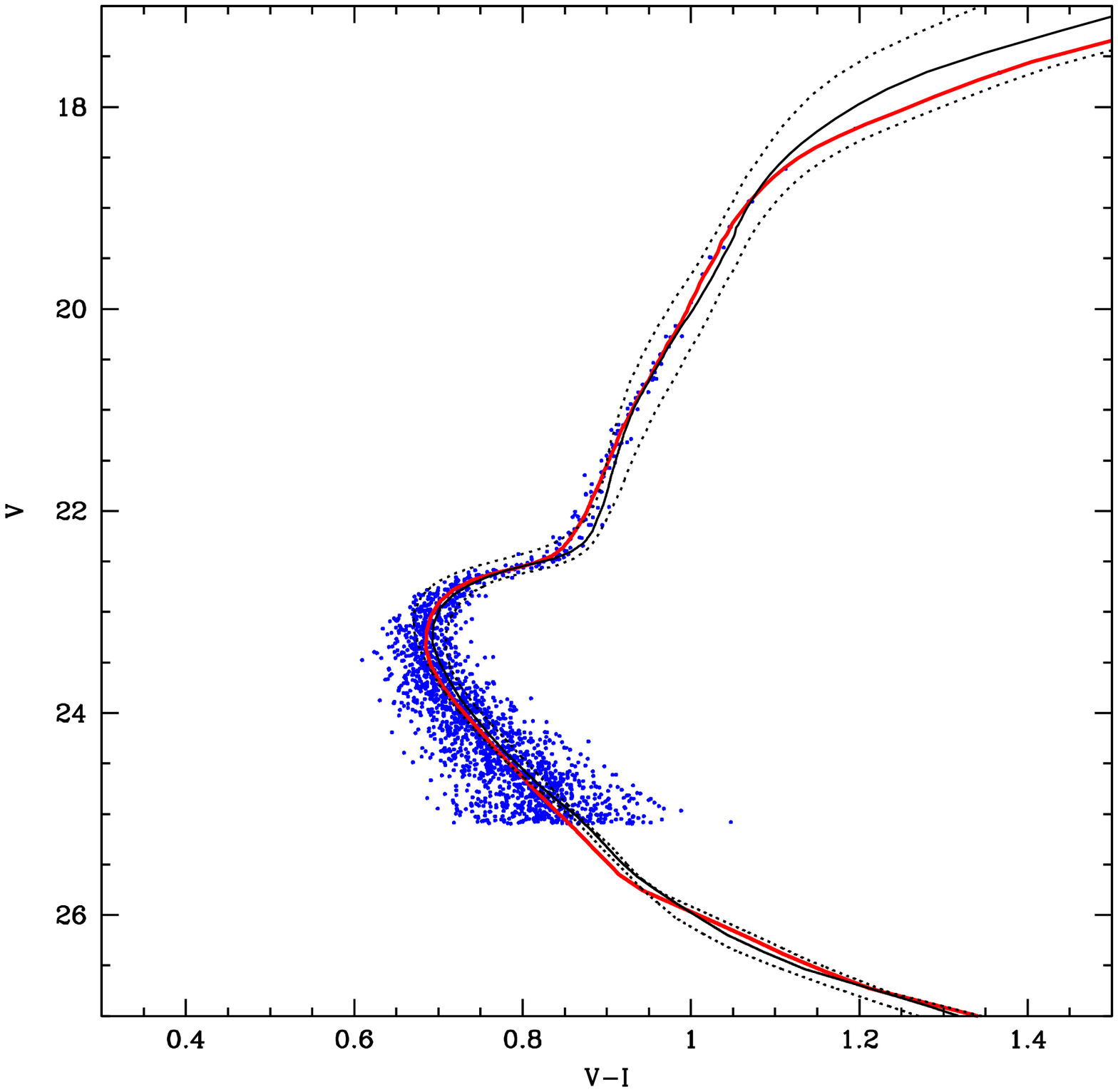}{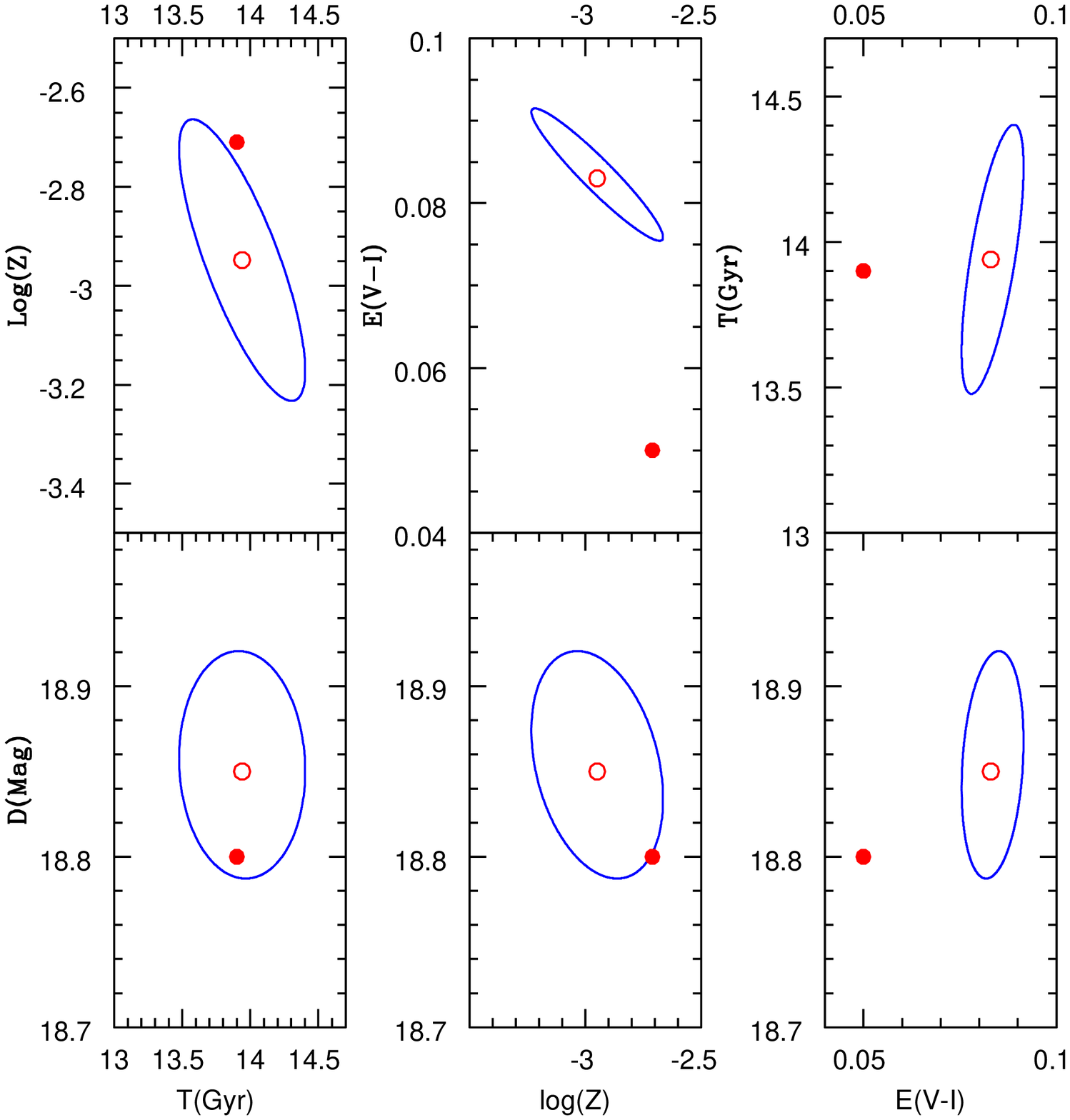}{Simulated Single Stellar Population CMD having an age of 13.9 Gyr and 
$\log Z=-2.7$. The thick (black) curve shows the input isochrone, with the 
optimal fit isochrone recovered by the method appearing as a thin (red) 
curve. Dotted curves show the youngest and oldest isochrones allowed by
the method, at the $1-\sigma$ level, which in fact span a 1.0 Gyr interval.}{The six panels show the projection of the error 
ellipsoid resulting from the Monte Carlo
simulation and inversion of a SSP with age 13.9 Gyr and metallicity
$\log Z=-2.7$, onto different planes. The filled circles show input parameters, and
the empty ones the average values for the recovered parameters.}{fig01}

We shall refer to $G_{i}(T,Z,D,R)$ as the
likelihood matrix, since each element represents the probability that a given star, $i$,
was actually formed with metallicity $Z$, at time $T$ with any mass, and that it was
observed with a distance modulus of $D$, and reddening of $R$. That is, in this Bayesian 
approach, we
treat the stellar mass $m$ as a nuisance parameter which we marginalize over. 

The probability that an entire observed CMD with $N_{\star}$ stars 
arose from a particular model, {\sl i.e.}, from a particular
choice of parameters $(T,Z,D,R)$, will now be given by:

\begin{equation}
{\cal L}(T,Z,D,R) \; = \; \prod_{i=1}^{N_{\star}} G_{i}(T,Z,D,R) \qquad .
\label{eq:merit}
\end{equation}

In this way, we have constructed an optimal merit function which can be evaluated
at any point within the relevant four dimensional parameter space $(T,Z,D,R)$, given
an observed CMD. It is important to remark that this merit function uses all information
available in the CMD densely. No binning or averaging over discrete regions is necessary, and no
aspects of the distribution of stars on the CMD are discarded. Also, all the rich 
stellar evolutionary physics is included in the comparison with the models, as the
isochrones are not being reduced to a discrete set of numbers (e.g. the position of the TO, RGB slope, etc.) 
or even curves on the CMD. 
As we shall see in section \S 3, the full use of all features in both data and models,
allows a precise estimation of the inferred parameters.

Following
Bayes theorem, one would now want to identify that point which maximizes \like, 
the 4 coordinates which maximize the probability that the observed CMD
 did in fact arise from a particular point in the parameter space, with
the set of physical cluster parameters which have the highest probability of
having given rise to the observed CMD. 

Several methods exist for identifying the global maximum of a multi dimensional surface. 
Most standard methods run into considerable difficulties in cases where local
maxima exist, as our problem turns out to have. The most reliable way of solving this
problem is to evaluate \like~ throughout a dense grid in all of the available
parameter space. This however, is highly time consuming, and hence it is highly 
desirable to find a more efficient method. 

We have used the genetic algorithm approach, where a number of ``agents''
are placed at random points within our parameter space, and the position of each codified
into a binary number, the ``genes'' of each agent. The number of binary bits used
for each coordinate fixes the resolution of the algorithm, which in this case corresponds 
to the resolution of the isochrone grid on the ($T,Z$) plane, and to 0.01 mag in the ($D,R$)
plane. The value of \like ~is then evaluated
at the position of each agent, and the result is used as a fitness function. A second 
generation is then constructed by ``mating'' the original agents, pairing them and combining
the ``genes'' of each parent to produce two ``offsprings''. The fitness function is used
to weight the mating probability, such that the fitter individuals have a greater
probability of ``mating''. The ``genes'' of the offsprings are a combination of those of their
parents, and hence as generations proceed, the population gradually moves into regions
where the fitness function is higher, {\sl i.e.}, into regions of high \like. To avoid getting
trapped into local maxima, a slight ``mutation rate'' is introduced, through which some of the
offsprings in each generation have a randomly selected ``gene'' switched from ``1'' to ``0'',
or vice-versa. This simplistic algorithm has been shown to converge remarkably efficiently
into the global maximum of complex multi-dimensional surfaces, associated to very different
problems. A more complete description of the algorithm can be found in Charbonneau (1995), with
successful applications to diverse astrophysical problems being described in {\sl e.g.} Teriaca et. al (1999) for 
fitting radiative transfer models to solar atmosphere data, Sevenster et al. (1999) in 
fitting parameters of Galactic structure models to observations, Metcalfe (2003), for the fitting of 
white dwarf astroseismology parameters or Georgiev \& Hernandez (2005) in the problem of
inferring optimal wind parameters from observed spectral line profiles, among many others.

\section{Testing the method}

In this section we test the method described in Section \S 2, through the inversion of
a series of synthetic CMDs. As in this case the point in parameter space from which 
a particular CMD was simulated is known, we can optimally asses the performance of the method.

In simulating an observed CMD the first step is to obtain an adequate isochrone library.
Different groups have succeeded in producing state-of-the-art stellar evolutionary codes
using mostly, but not all the time, the same input physics. Since systematic errors
will be produced by the choice of the set of isochrones, we use the results of the
three different groups (Bergbusch \& VandenBerg 1997, VandenBerg et al. 2006, 
Demarque et al. 2004, Girardi et al. 2002).
  We have worked directly with the output of stellar evolutionary codes
to produce our own  isochrone grid with a dense grid including
 180 time steps and 120 metallicity intervals. Each 
isochrone contains approximately 300 stellar masses. The time resolution for ages of 
between 4.5 Gyr
and the upper limit of 18.0 Gyr is of 0.2 Gyr, with younger ages being sampled more densely. 
The spacing in metallicity is linear in the logarithm, with 0.03 dex per interval. Throughout the paper
we shall identify metallicity with [Fe/H], when using theoretical isochrones, or comparing against
empirical inferences. The spacings
of the isochrone grid will determine the maximum resolution of the method, independently of
what approach is taken at the following levels of the problem. For the purposes of 
presenting the method, these grids are more than appropriate. 

It is important to note that 
the highly non-linear variations in magnitude and colour with mass which appear along isochrones
on critical regions of the CMD implies that a direct interpolation between two isochrones will
not produce the intermediary one, but an average curve which bares little resemblance to the
output of a stellar evolutionary code.  In all cases, we have limited
the isochrones to below the helium flash. This condition restricts the information we consider
in our method to the region where disagreement between different groups working in the field is 
at a minimum.

Once the isochrones have been established, an IMF is chosen to distribute stars amongst different
masses. This function proves to be relatively unimportant, as the mass range of stars one is dealing
with is small provided stars are selected from one magnitude or so below the TO region all the way to the
tip of the RGB. This ensures that the mass range probed by the observations is small. For
example, in globular clusters older than $\sim 5$ Gyr the mass interval is around $1 M_{\odot}$
and hence in this narrow mass range the details of the IMF are irrelevant.  
The distribution
of stars along the isochrone is a much more sensitive function of the stellar evolutionary models, 
which is in fact what determines that the main sequence is highly populated, the red giant branch much more
sparsely so, and that the HR gap region will contain very few stars. The IMF we use is that of
Kroupa et al. (1993), although as mentioned above, any other approximation to this function would yield the same
results (see \S4.1).

\figce{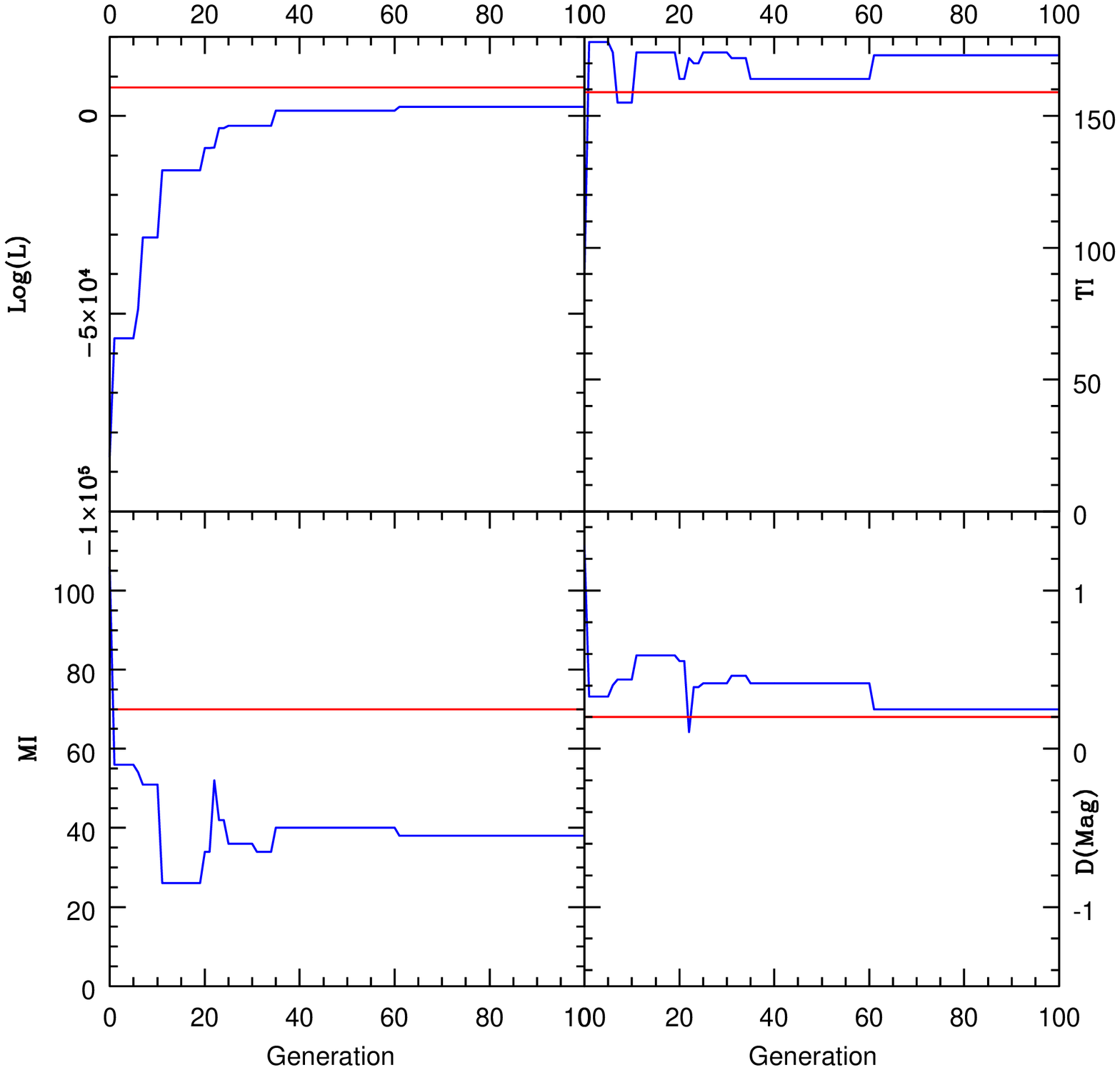}{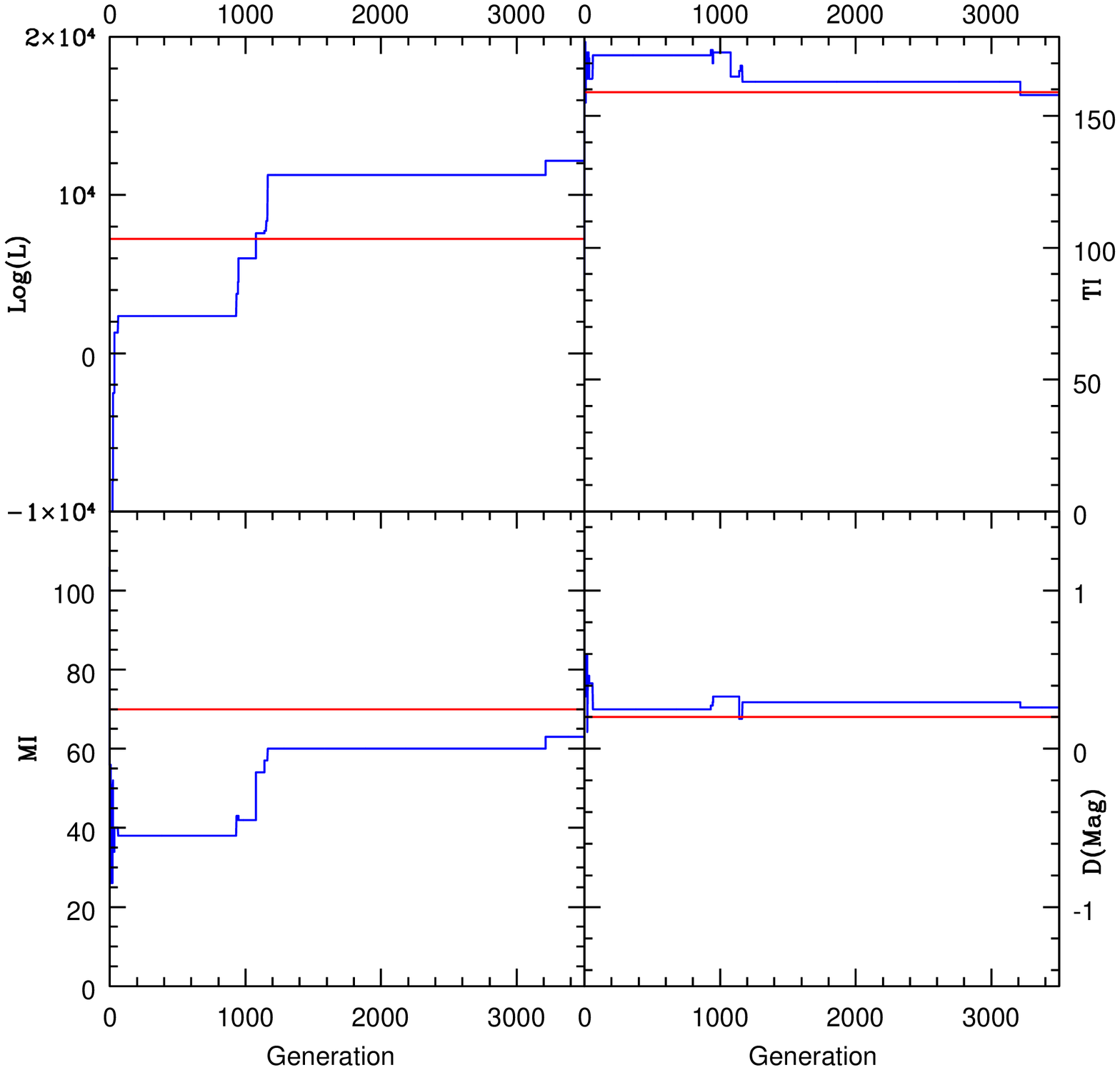}{The four panels show, in a clockwise progression starting at the upper left,
the maximum likelihood point found by the method, the optimal time index,
the optimal distance modulus, and the optimal
metallicity index, all as functions of generation number for the genetic algorithm
simulation, for the first 100 generations.}{Same as the left panels, 
but showing the full 3500 generations of the genetic algorithm simulation.}{fig02}

Having an isochrone grid and an IMF, we can now pick a point in our $(T,Z,D,R)$ parameter space 
and place a chosen number of stars into the magnitude-colour plane. For the first implementation
we have simulated an extreme example with close to 2000 stars around point $(13.9, -2.7, 18.8, 0.05)$, a single stellar
population of age 13.9 Gyr, metallicity $-$2.7 (that is, close to $-$1.0 
in solar units, for a solar metallicity of $\log Z_{\sun}=-$1.7), 
with a distance modulus of 18.8 magnitudes, and observed through a reddening of 0.05 magnitudes. Next we model 
magnitude and colour errors as Gaussian with sigmas as appropriate for current HST or high quality long 
integration ground-based observations.  Explicitly we used 
$$
\sigma(L)=0.001 \times \exp\left[ (L+D-17.0) / 2.25 \right] 
$$
$$
\sigma(C)=0.001 \times \exp\left[ (L+D-17.0) / 1.97 \right]		
$$
where $L$ is the magnitude, in the $V$ band, of the theoretical star, and $C$ its
$V-I$ colour. At this point we can now produce a synthetic CMD, and for this first
test obtain the 2009 points shown in Figure \ref{fig01}. A magnitude cut at 
$V=25$ has been used, limiting analysis to stars brighter than this cutoff. Given the highly 
correlated way in which star populate the main sequence, together with growing
and sometimes divergent errors towards higher magnitudes, it is sufficient to 
restrict analysis to 2 magnitudes below the 'turn off' region. In what follows
we shall keep the 25 magnitude lower limit, although a changing magnitude cut
as described above yields the same results.

At this point, Equation \ref{eq:merit} can be used to obtain the likelihood function of the
simulated CMD with respect to any given point on our 4-D parameter space. As discussed in
section \S 2, the seeking of the optimal point will be carried out through a genetic algorithm simulation,
the workings of which, for this first example, we describe below.

The upper left panel of Figure \ref{fig02} shows the evolution of the value of the likelihood 
merit function of Equation \ref{eq:merit} over the first 100 generations of the genetic
simulation. The horizontal line gives the value of this same function, evaluated
at the actual point in our $(T,Z,D,R)$ parameter space which was used to produce
the synthetic CMD being inverted. This first panel shows an increase with the generation
number of the optimal value of the likelihood function, as the absolute best point
found in all the simulation up to a given generation is what is being plotted. 
This value increases very rapidly at first, as the initially random points begin to
migrate towards regions more  closely resembling the simulated CMD, but progress
becomes increasingly slower. In fact, there is no definitive criterion to establish
when a genetic algorithm simulation has finally converged, this has to be inferred from
simulations with synthetic data and repeated experiments, as an approximate number of
generations, suitable for a particular problem, after which no further improvement
is achieved. For the problem being treated here, we have found the method stops
yielding any improvement in the fits in general well before 3500 generations. 
We have thus ended the genetic simulations at 3500 generations.
The value of the likelihood function evaluated at the actual input point always remains
above the one corresponding to the optimal one found by the simulation for this first
100 generations shown here.

The upper right panel of Figure \ref{fig02} gives the evolution of the
optimal time grid index (TI) along the isochrone grid, found up to a certain point.
In this case, the input age of 13.9 Gyr corresponds to our time index 159 out of 180,
over this initial 100 generation period the time index corresponding to the optimal
model found fluctuates from an initial value of 120 to 178, from 6 to 18 Gyr, and ends
this initial period 12 grid points above the input value, at 16.9 Gyr. The bottom left and right
panels show the corresponding evolution of the metallicity index and the distance modulus, 
again, the horizontal line gives the input parameters. We can see that after 2 generations
the initially random values have approached significantly the input ones, and then 
show considerable variations
of around 35 metallicity grid intervals (1.5 dex) and 0.8 mag in the distance modulus.

\figce{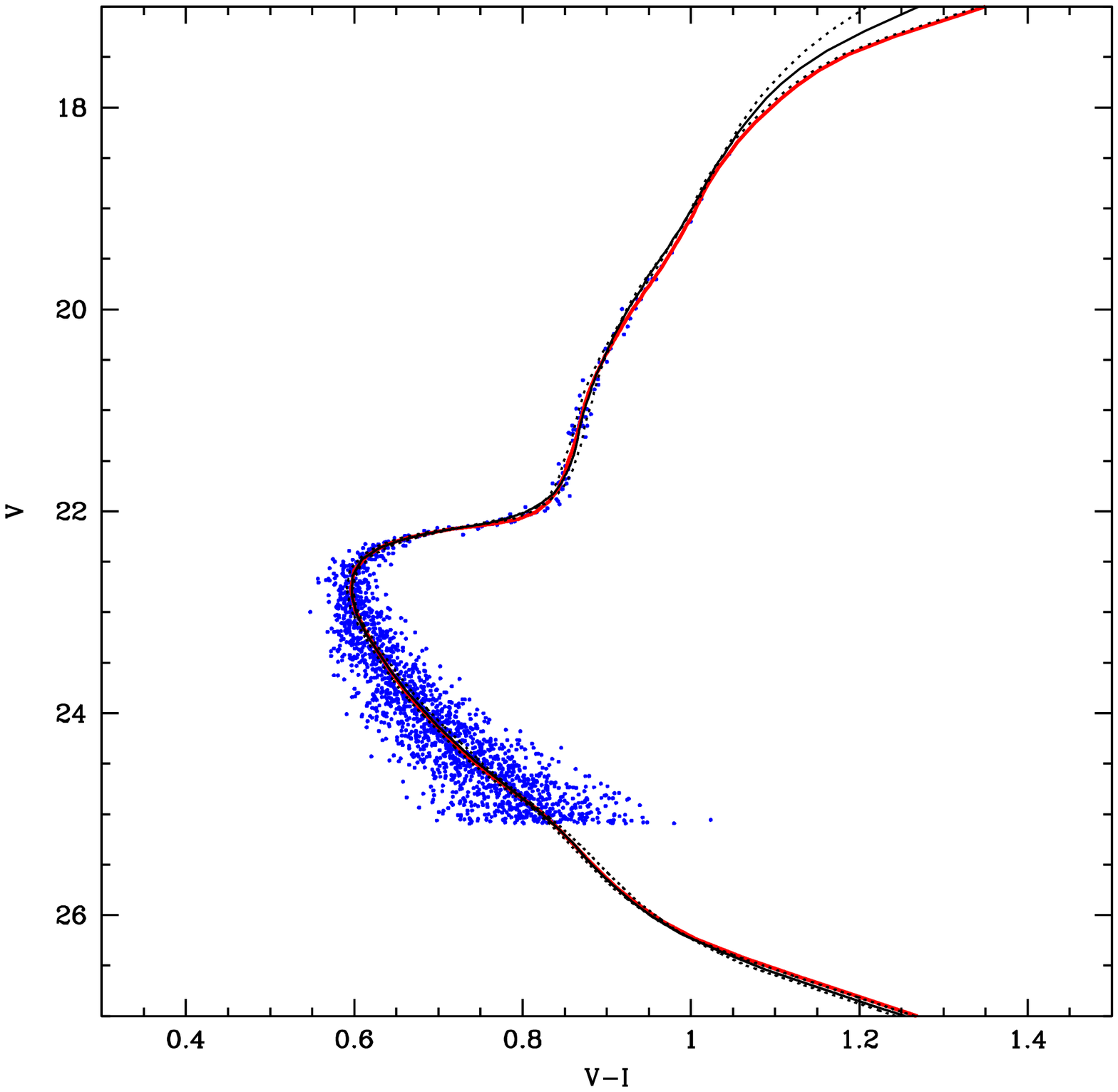}{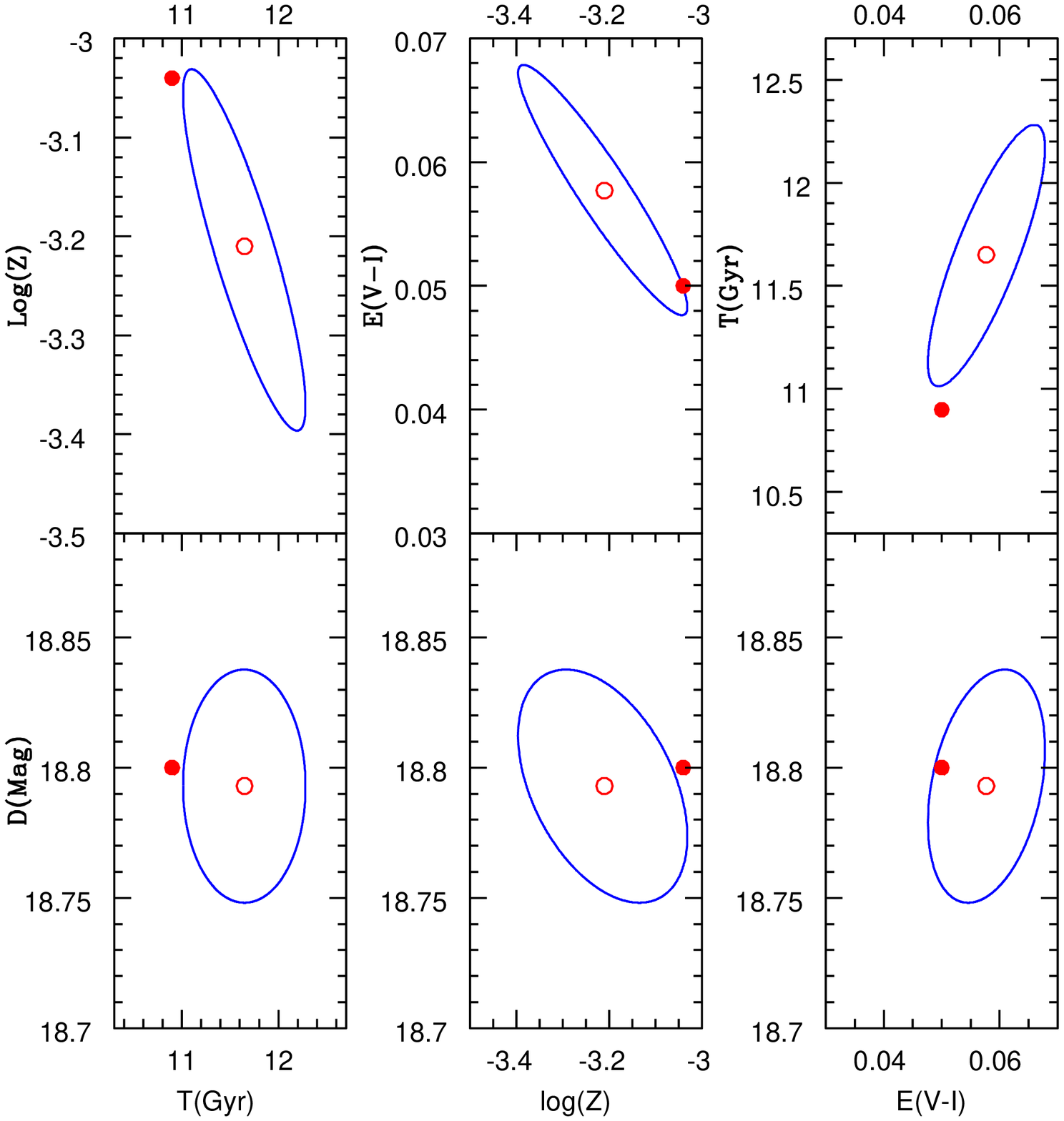}{Simulated Single Stellar Population CMD having an age of 10.9 Gyr and 
$\log Z=-3.05$. The thick (black) curve shows the input isochrone, with the 
optimal fit isochrone recovered by the method appearing as a thin (red) 
curve. Dotted curves show the youngest and oldest isochrones allowed by
the method, at the $1\sigma$ level, which in fact span a 1.48 Gyr interval.}{The six panels show 
the projection of the error ellipsoid resulting from the Monte Carlo
simulation and inversion of a single stellar population with age 10.9 Gyr and metallicity
$\log Z=-3.05$, onto different planes. The filled circles show input parameters, and
the empty ones the average values for the recovered parameters.}{fig03}

It is interesting to re-plot the left panel of Figure \ref{fig02} this time over the entire 3500 generation range, 
this is done in the right panel of Figure \ref{fig02} where we can see that substantial improvements in the 
optimal likelihood value found are still occurring after 1000 generations. A striking feature seen in the upper
left panel of this last Figure is that
the genetic algorithm has actually found an optimal model which in fact yields a likelihood value
above that of the input parameters. At around 1000 generations, the optimal likelihood
found crosses the horizontal line giving the likelihood value of Equation \ref{eq:merit} of the input parameters,
with respect to this particular realization of the CMD corresponding to them. This apparently
paradoxical result is in fact unavoidable, and highlights the statistical nature of the problem.
The construction of a CMD from a point in the single stellar population parameter space can not
be thought of as neither a deterministic nor a unique process. Even before the errors associated with
procuring the observations enter the picture, the process of sampling an IMF has already introduced
a probabilistic ingredient intrinsic to the cluster itself ({\sl e.g.} Cervi\~no et al. 2002, 
Cervi\~no \& Valls-Gabaud 2003). In this sense, obtaining a single 
simulated CMD and using it to compare with an observed one can only be at best a crude
first approximation to the inverting of a CMD. In this particular case, the convolving of the 
theoretical stars with error gaussians has shifted the points around in such a way that the isochrone
(understood as a series of stellar evolutionary models, not just a curve on a CMD) which
has the greatest probability of having yielded the CMD is not the input one, but one which is one
time grid interval above (0.2 Gyr), 11 metallicity grid intervals below (0.4 dex) and at a distance modulus 
around 0.05 magnitudes larger than the input parameters.

The above result makes it clear that inferring structural parameters of stellar populations, even in the
simplest cases, can not be attempted without a full consideration of the probabilistic nature of the
problem, and also, can not give error-free results. However, it is possible to evaluate the intrinsic
errors of the method, in connection to any particular implementation, by the Monte Carlo method.
The experiment is repeated a large number of times, and the various answers analyzed to obtain
mean inferred values with well determined dispersions and correlations.

This method naturally takes into consideration all internal sources of error and dispersion
inherent to the procedure being tested, and yields accurate and reliable confidence intervals on 
the inferences obtained. Figure \ref{fig01} (right) shows with filled circles the input parameters
of the CMD of Figure \ref{fig01} and with empty ones the center of the distributions of the
recovered parameters after the whole procedure of constructing a synthetic CMD and
inverting it was repeated 20 times. The ellipses give projections of the error ellipsoid
onto the different planes being shown. It can be seen that the only systematic which
appears in this case beyond the $1\sigma$ level is an $+$0.03 magnitude offset in the 
recovered values of the reddening for the cluster. All other parameters have been recovered 
with no systematics, the $1\sigma$ errors for this 13.9 Gyr old cluster being 
0.5 Gyr in age, 0.38 dex in metallicity and 0.05 mag in the distance modulus.

The isochrone corresponding to the solid circles is shown in Figure \ref{fig01} by the thick solid curve,
which practically lies above the thin (red) curve of the isochrone corresponding to the 
empty circles along all the CMD. Some differences between the two isochrones are evident
along the upper half of the red giant branch, where so few stars are expected, that
none of the 2009 obtained in the CMD realization shown appear there. The dotted curves give the
isochrones corresponding to the oldest and youngest points on the ellipse in the
age-metallicity plane, which are seen to bracket the HR-gap and RGB regions of the 
CMD shown.

The second example was constructed from point $(10.9, -3.05, 18.8, 0.05)$ in our parameter
space, taken to represent a typical old Single Stellar Population (SSP). In testing the method for 
systematics related to effects not explicitly included in the model, comparisons will be made
at this point in parameter space, in section \S~4.

Figure \ref{fig03} shows the CMD
which results from convolving the modeled stars with the same simulated errors than last time.
Also shown are the isochrone from which the simulated stars were constructed, thick (black) curve,
and the optimal one found by the genetic algorithm, thin continuous curve. The dotted curves give
the oldest and youngest isochrones consistent with the Monte Carlo simulation on the inverting 
procedure at a $1\sigma$ level, this time spanning 1.2 Gyr. 
Analogous to Figure \ref{fig01}, Figure~\ref{fig03}(right)
gives the projection of the error ellipsoid for the inferred parameters using the Monte Carlo method,
projected onto 6 different planes. One again sees that there are no systematics beyond the $1\sigma$ level,
from which we conclude that the method accurately recovers the input parameters of the cluster, and yields
meaningful confidence intervals for the answer supplied.

\figce{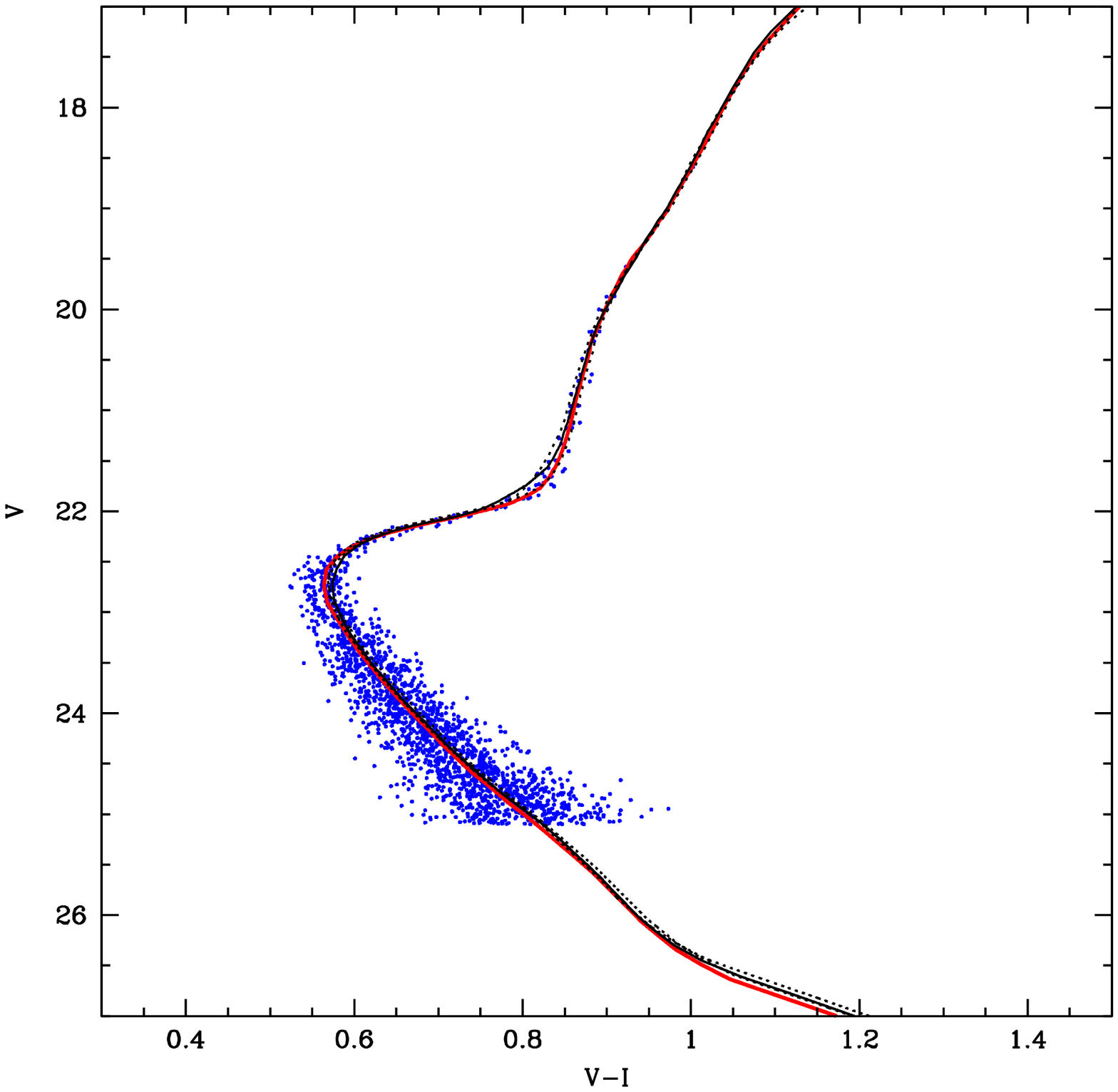}{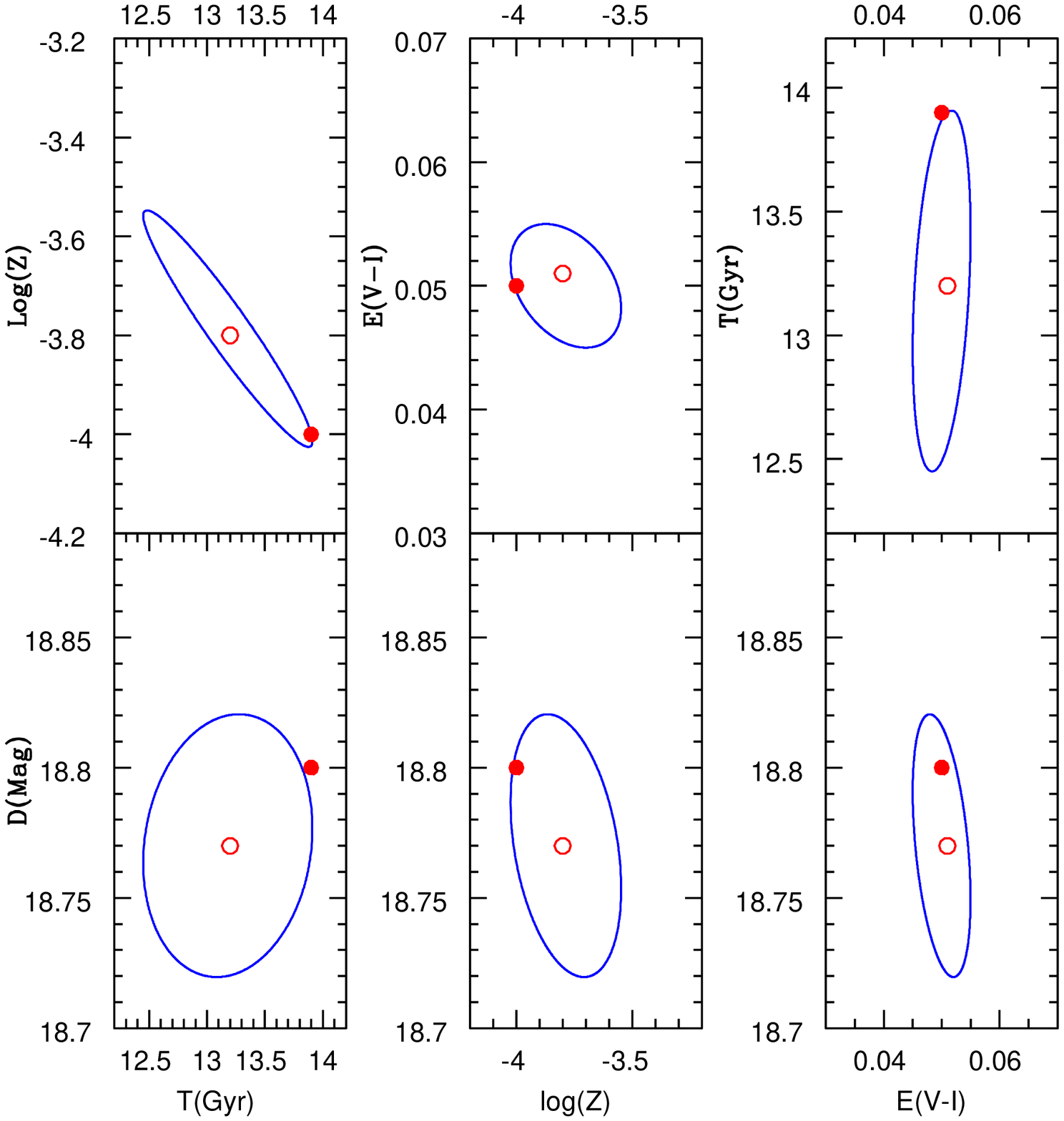}{Simulated Single Stellar Population (SSP) CMD having an age of 13.9 Gyr and 
$\log Z=-4.0$. The thick (black) curve shows the input isochrone, with the 
optimal fit isochrone recovered by the method appearing as a thin (red) 
curve. Dotted curves show the youngest and oldest isochrones allowed by
the method, at the $1\sigma$ level, which in fact span a 1.44 Gyr interval.}{The six panels show the projection of the 
error ellipsoid resulting from the Monte Carlo
simulation and inversion of a SSP with age 13.9 Gyr and metallicity
$\log Z=-4.0$, onto different planes. The filled circles show input parameters, and
the empty ones the average values for the recovered parameters.}{fig04}

Looking at the upper left panel of Figure \ref{fig03} (right) 
we clearly see the actual intrinsic  age-metallicity degeneracy 
in the way the  $1\sigma$ ellipse in the
($T,Z$) plane slants from the upper left to the lower right. Optimal fits with large ages tend to
correspond to lower metallicities, and vice-versa. However, we see that having included all the
information available both in the simulated CMD and in the theoretical isochrones minimizes this
degeneracy and one obtains closed confidence intervals. If the CMD is reduced to, say two numbers,
corresponding to the location of the 'turn off', when comparing to a similarly restrictive view of
what the stellar evolutionary models encode, one ends up with a confidence region which spans an extensive
region of the ($T,Z$) plane, merely a relation between the possible age and metallicity, with no way of 
restricting the answer to a narrower region. The above highlights problems inherent to methods where
information is being discarded, to which must be added ambiguities associated to the definition of the
'turn off' point. Given the current high accuracy of the observations and the theoretical
isochrones, inspection of simulated and observed CMDs makes it clear that there is in fact no such thing as a 
turn off 'point', but rather a diffuse, hazy region which is prone to sampling
fluctuations and the presence, or otherwise, of a few odd stars.

Other correlations are evident in Figures \ref{fig01}  and \ref{fig03}, for example a  clear negative
correlation between the inferred reddening and metallicity, also a generic feature
of any stellar population inversion method, inherent to the structure of the isochrones.
It is interesting to note in the lower 3 panels of Figures \ref{fig01} and \ref{fig03} that the inferred
distance to the cluster shows no strong correlations with any of the other 3 parameters,
contrary to what happens when the isochrones are reduced to a set of numbers such as
location of the turn off point, slope of the RGB etc.

For a third example, in trying to cover points on our parameter space representative of
old SSPs, we simulated an old, metal poor such system. Starting from
point $(13.9, -4.0, 18.8, 0.05)$ we construct a series of CMDs, 
one of which is shown in Figure 4,
with curves analogous to what appears in Figures 2 and 3. It is impressive to see how
nearly indistinguishable the shapes of the input and mean recovered isochrones are on the CMD, together
with the isochrones corresponding to the oldest and youngest isochrone allowed by the Monte Carlo simulation at
a  $1\sigma$ level. All 4 of these curves appear practically on top of each other in this case, with
very minor differences only apparent  at the curvature between the HR gap region and the RGB.

The above is particularly surprising when considering that the isochrones shown span almost
a 1.5 Gyr-long time interval. Of course, the other parameters besides the age have been adjusted to
maximize the fit when shifting the age, in accordance with the correlations shown in Figure \ref{fig04},
the oldest isochrone has the lowest metallicity together with a somewhat above average reddening
correction. This point again shows that the isochrones have to be treated as collections of full
stellar evolution models, not reduced to a few scalar shape parameters, not even to simply shapes on the
CMD. Having used the full shape of the isochrones continuously throughout the full CMD, together
with information regarding the duration of different stellar phases in constructing the merit function
of Equation \ref{eq:merit} has allowed a much finer discrimination within our  $(T,Z,D,R)$ parameter space, 
much beyond what considering only the shape of the isochrones would have allowed. 
Changing the input parameters of the isochrone even at a 2-3 sigma level, would result in curves compatible
with the shape of the simulated stellar distribution, but which are ruled out due to having included all
available information both in the CMD and the stellar models.

\figce{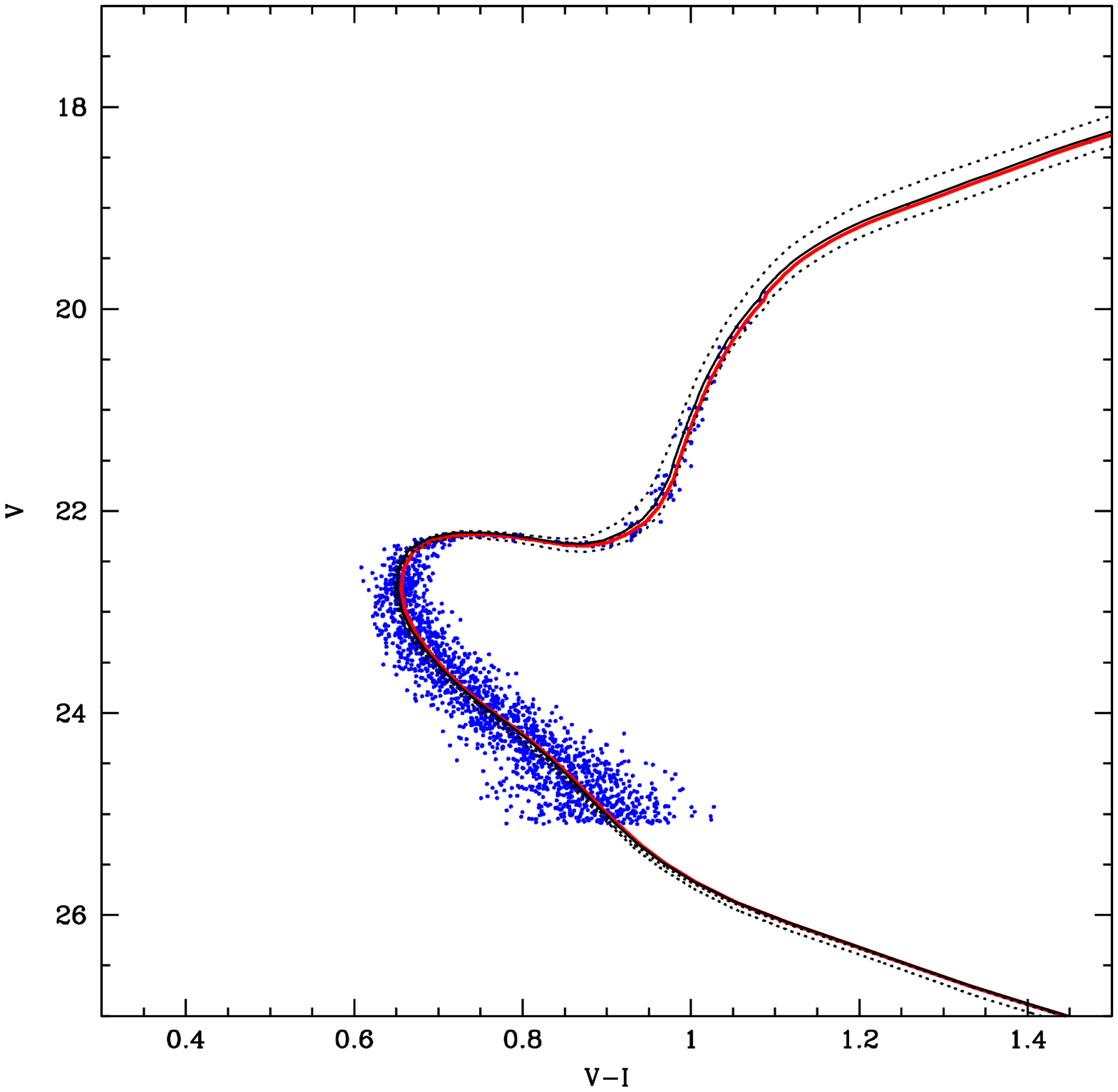}{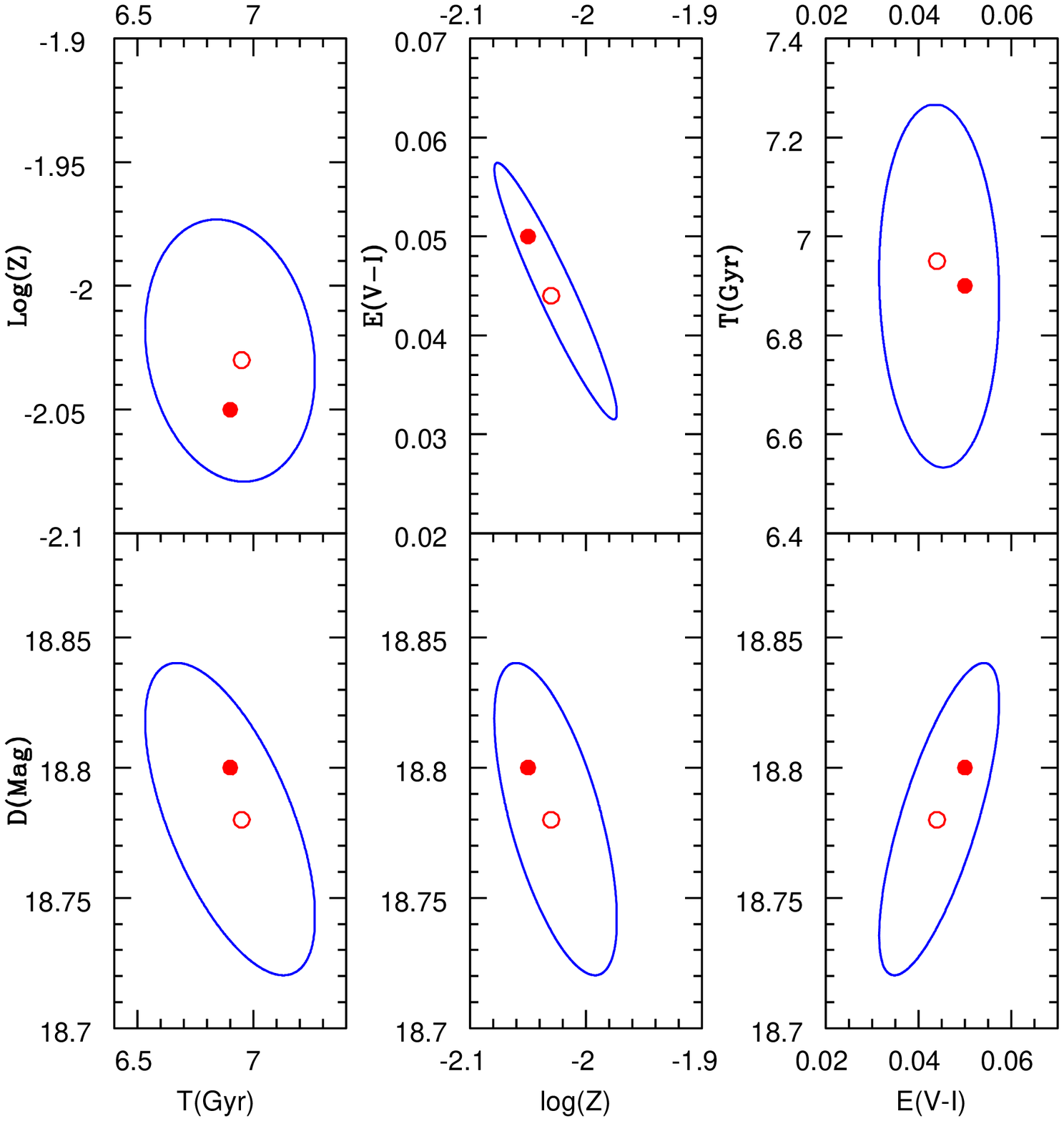}{Simulated Single Stellar Population CMD having an age of 6.9 Gyr and 
$\log Z=-2.05$. The thick (black) curve shows the input isochrone, with the 
optimal fit isochrone recovered by the method appearing as a thin (red) 
curve. Dotted curves show the youngest and oldest isochrones allowed by
the method, at the $1\sigma$ level, which in fact span a 1.0 Gyr interval.}{The six panels show the projection of the error 
ellipsoid resulting from the Monte Carlo
simulation and inversion of a SSP with age 6.9 Gyr and metallicity
$\log Z=-2.05$, onto different planes. The filled circles show input parameters, and
the empty ones the average values for the recovered parameters.}{fig05}

Again, no systematic
offsets between the input values and the mean recovered ones appears beyond the  $1\sigma$ level,
as can be seen from Figure \ref{fig04}. In this same Figure we see the same correlations evident in the
previous analogous Figures, with the confidence intervals, particularly in age and metallicity
having grown considerably in comparison to Figure \ref{fig03}. This is obvious when one considers
how consecutive isochrones equally spaced in time will show progressively smaller differences
as the age of a stellar population increases. Having left the assumed observational errors and
stellar numbers constant,
the uncertainties in the inferred parameters have grown in going to an old, metal poor 
population.

For the fourth and fifth examples we complete our sampling of SSPs 
parameter space with two clusters at $(6.9, -2.0, 18.8, 0.05)$ and  $(6.9, -3.37, 18.8, 0.05)$.
Figures \ref{fig05} and \ref{fig06} are completely analogous to the corresponding previous
figures and essentially illustrate the same points mentioned before. 
For these younger simulated stellar populations the mean recovered parameters appear within
one grid interval in metallicity of the input parameters, the theoretical
maximum resolution of the implementation has been realized at these young ages, even 
after having considered realistic errors in magnitudes and colours. The age resolution
is still above that of the isochrones, but robust $1\sigma$ errors of around 0.5 Gyr 
are encouraging.

We add that the precision
of the method in recovering distance modulus and reddening parameters is sufficient to
make it a useful tool in the determination of distances to single stellar populations, with errors 
sometimes as good as a few hundredths of a magnitude, as seen in Figure \ref{fig05}.
Again, the correlations persist, although the details of these correlations change
from example to example in ways that reflect the details of the isochrones around each region, and 
which are dependent on both the magnitude of the assumed observational errors and the actual 
numbers of stars present in the simulated CMDs. 

With simulated errors tending to zero the inferred parameters would tend to the input ones,
to within the resolution of the isochrone grid used, always. Also, if the statistical
comparison between data and models is robust, the systematics and confidence intervals
should tend to zero as the number of data points tends to infinity, which of course is
never the case.

\section{Systematics}

Having tested successfully the method across representative points in the $(T,Z,D,R)$ space, 
we now turn to the problem of estimating how robust the inferences of our method will
be in more realistic cases, where physical ingredients not explicitly considered in the 
statistical modeling behind Equation \ref{eq:merit}, and present in a real application, will 
complicate matters, and might result in systematic offsets between the actual physical
parameters of a stellar population and our inferences.

In this section we explore such cases by constructing synthetic CMDs under assumptions
different to the ones the method uses in the inversion procedure, and compare
how serious these effects are in altering our inferences. The 5 effects considered will be:
(1) Errors in the assumed IMF, (2) the presence of unresolved binaries, 
(3) the effects of reducing the numbers
of stars present in the observed CMD, (4) systematics associated with the different sets of stellar tracks
used in constructing the isochrone grids, and (5) the presence of contaminating stars in the
observed CMDs.

\figce{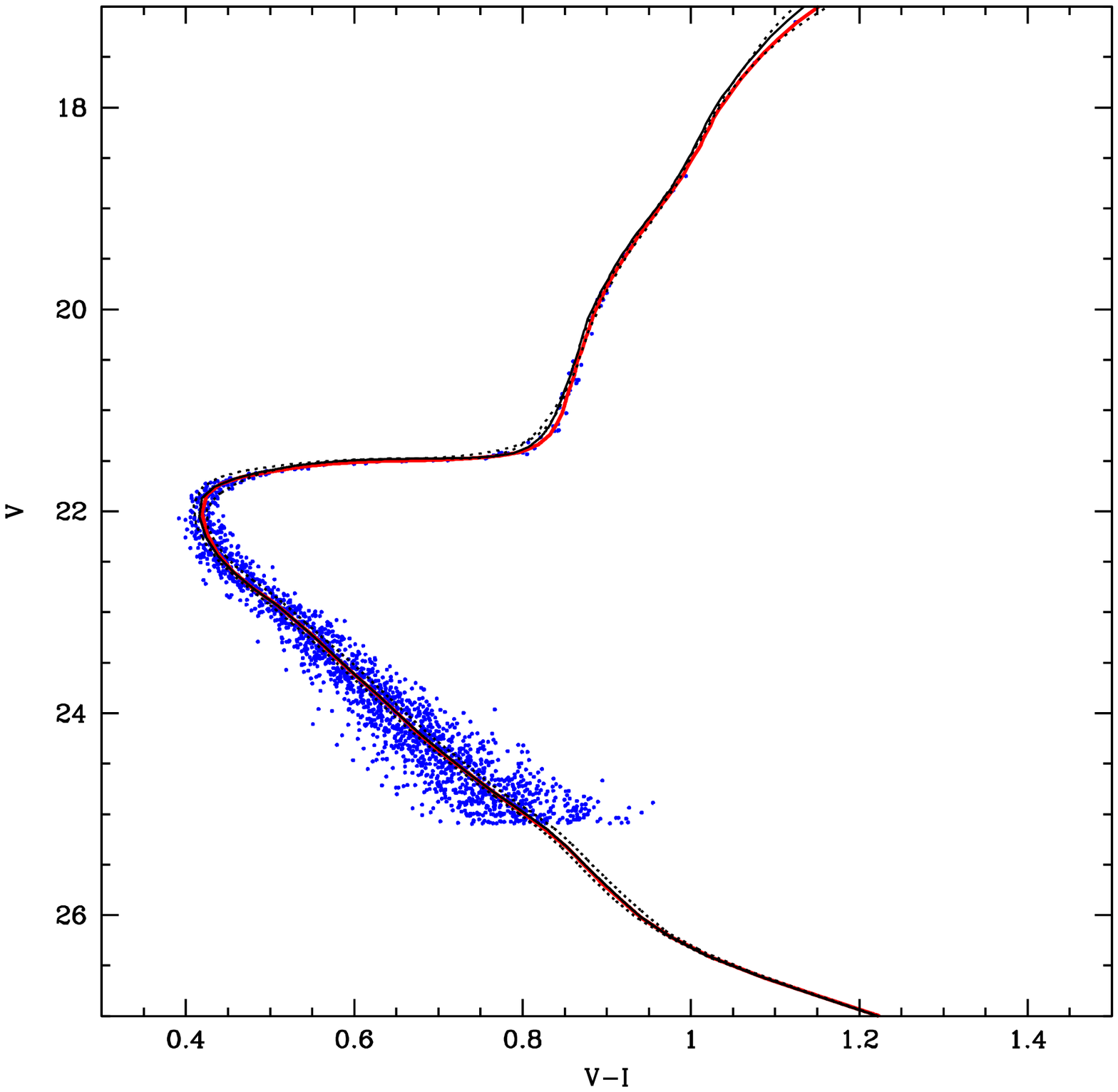}{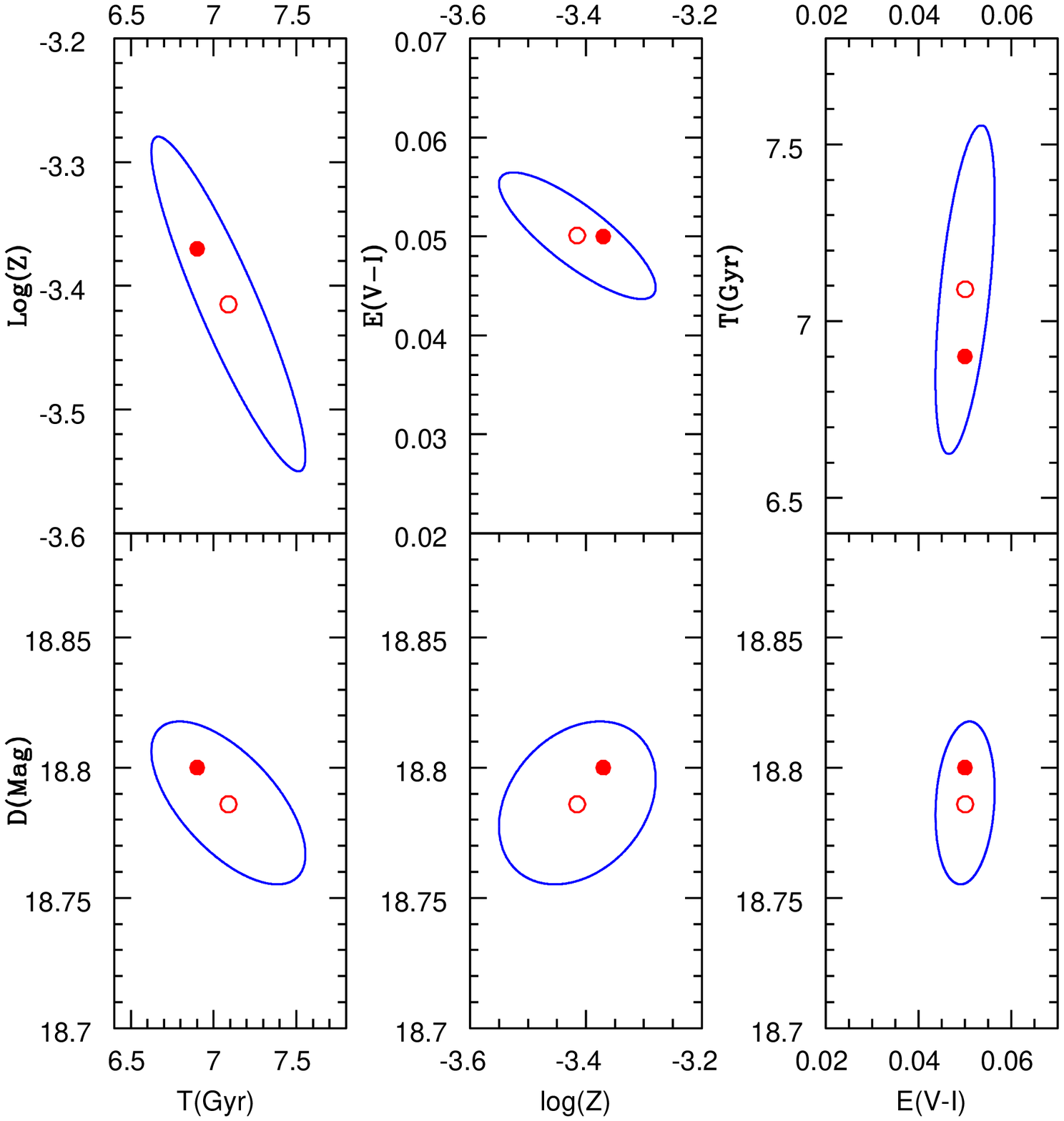}{Simulated Single Stellar Population CMD having an age of 6.9 Gyr and 
$\log Z=-3.37$. The thick (black) curve shows the input isochrone, with the 
optimal fit isochrone recovered by the method appearing as a thin (red) 
curve. Dotted curves show the youngest and oldest isochrones allowed by
the method, at a $1\sigma$ level, which in fact span a 1.0 Gyr interval.}{The six panels show the projection of
 the error ellipsoid resulting from the Monte Carlo
simulation and inversion of a SSP with age 6.9 Gyr and metallicity
$\log Z=-3.37$, onto different planes. The filled circles show input parameters, and
the empty ones the average values for the recovered parameters.}{fig06}

\subsection{Effects of the IMF}

Figure \ref{fig07} gives a CMD resulting from the same point in parameter space as
Figure \ref{fig03}, $(10.9, -3.05, 18.8, 0.05)$, but created this time using a substantially different IMF, 
heavily weighted towards  massive stars, increasing the slope with respect to the Kroupa et al. (1993)
IMF such that the typical stellar mass increases by a factor of 3. 
Obtaining the same number of points thus requires
turning a much larger total mass into stars, but under the assumption of an equal number
of stars as in Figure \ref{fig03}, the results are highly similar. The precise manner in which
the density of stars varies along the main sequence region is somewhat different from 
the case of Figure \ref{fig03}, but given the highly correlated manner in which magnitudes and
colours scale along this region, the information content of the CMD is only marginally altered.
The highly determinant sections beyond the main sequence span such a small mass range that
all of these features are virtually left unchanged by even the strongest modifications in the
IMF.

The inversion procedure was repeated
20 times, but assuming the same IMF as in Figure \ref{fig03}, which in this case no longer
corresponded to the IMF used to create the synthetic CMDs.
The results of the Monte Carlo procedure can be seen in Figure \ref{fig07},
which from the previous discussions would not have been expected to show any
substantial difference to Figure \ref{fig03}, as is indeed the case.  
By comparing with Figure \ref{fig03} we note no substantial differences, all inferred
parameters are well within the  $1\sigma$ error estimates, with very low uncertainties
of around 0.5 Gyr in age, 0.2 dex in metallicity, 0.01 and 0.05 mag in reddening
and distance modulus, respectively. The isochrones corresponding to the input, inferred,
and oldest and youngest values recovered (at a $1\sigma$ level), practically overlap
over the entire region over which significant numbers of stars are found. We see that
the theoretical expectations of the previous sections are borne out, and our results
are in fact highly robust to even large scale features of the IMF. This is reassuring from the 
point of view of inferring ages, metallicities, distances and reddening corrections of
resolved stellar populations, but disheartening from that of inferring the IMF of observed
stellar clusters through methods similar to what has been developed here.

\subsection{Effects of unresolved binary stars}

The second test was performed again using the same isochrone library, and the input point
 $(10.9, -3.05, 18.8, 0.05)$ of Figure \ref{fig03}, but this time to one in two of the simulated
stars we added an unresolved companion, chosen from the same IMF, {\sl i.e.}, the synthetic CMD was 
produced assuming a 50\% unresolved binary contribution. The inversion procedure was however exactly
the same as in the previous cases, with Equation \ref{eq:merit} remaining unchanged, a binary fraction of
0\% was assumed in recovering the inferred parameters.

One of the resulting CMDs which was
used in the Monte Carlo procedure for the inversion is shown in Figure \ref{fig08}, together with isochrones
as in the previous Figures. The presence of 50\% binaries is particularly conspicuous along
the main sequence, where the distribution of points is clearly smeared towards redder
colours, evident when
comparing, for example, with Figure \ref{fig07}. A few binaries can also be seen around the critical turn off
region, but these have had little effect. Given the large difference in luminosity between a RGB
star and any main sequence one (the most likely secondary to be added to a binary), binary
pollution along the RGB produces absolutely no effects. 

By looking at the best fit inferred isochrones at the base of the main sequence, it can be seen
that these have all shifted towards the right, in response to the weighting of the distribution
of points towards this direction. By looking at Figure \ref{fig08} one can see that this shift towards the
right has been achieved optimally by the genetic algorithm not by altering the reddening correction,
but rather through an offset in the distance modulus. Whilst inferred age, metallicity and reddening parameters
have been obtained within the $1\sigma$ error ellipses, the inferred distance modulus is off at 
around a $2\sigma$ level, which however corresponds only to 0.15 magnitudes. The clearest correlation
between the inferred parameters is still in the age-metallicity plane, and remains in the same sense
as expected. 

\figce{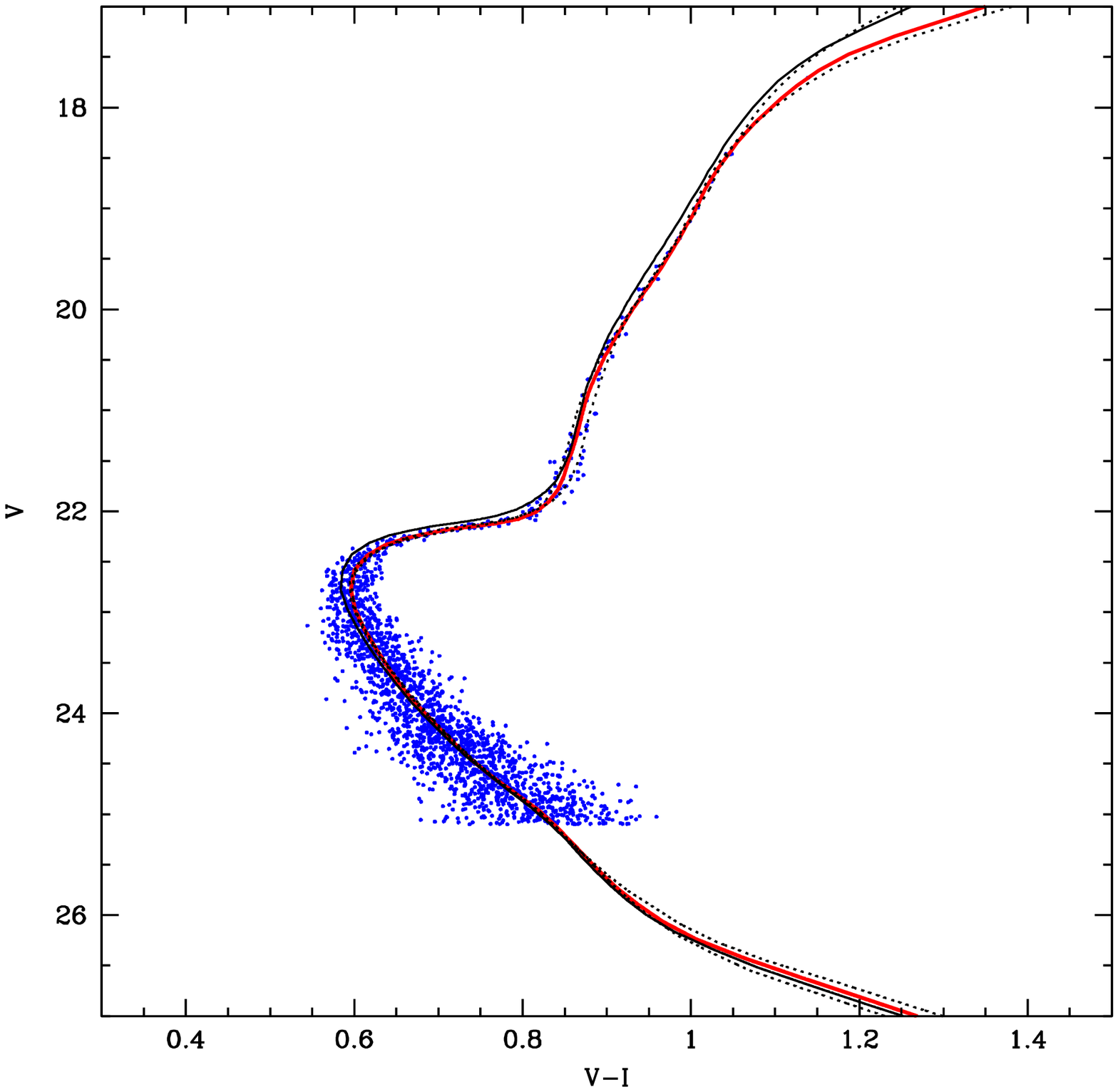}{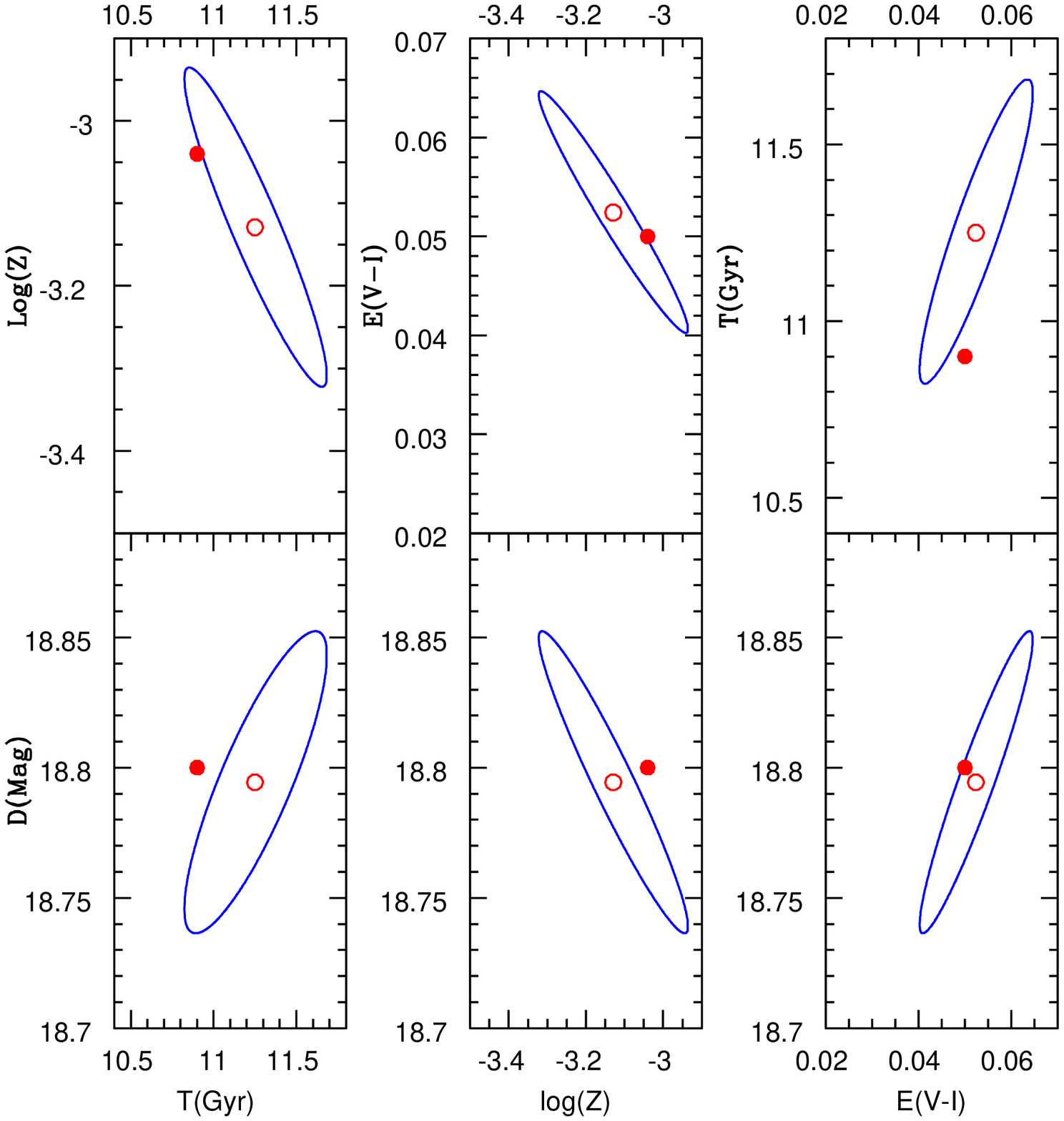}{Simulated Single Stellar Population CMD having an age of 10.9 Gyr and 
$\log Z=-3.05$. The thick (black) curve shows the input isochrone which was used assuming 
a high mass weighted IMF, with the 
optimal fit isochrone recovered by the method, assuming the standard IMF, appearing as a thin (red) 
curve. Dotted curves show the youngest and oldest isochrones allowed by
the method, at the $1\sigma$ level, which in fact span a 0.9 Gyr interval.}{The 
six panels show the projection of the error 
ellipsoid resulting from the Monte Carlo
simulation and inversion of a SSP with age 10.9 Gyr and metallicity
$\log Z=-3.05$, onto different planes, for CMDs constructed using a high mass weighted IMF
and inverted using the standard IMF. The filled circles show input parameters, and
the empty ones the average values for the recovered parameters.}{fig07}

Other than the slight offset in inferred distance modulus, the only other effect of having
introduced a relatively high binary population has been the broadening of the errors.
The confidence region is this time 2.5 Gyr across, compared to 1.4 Gyr for Figure \ref{fig03}. 
Again, the isochrones in Figure \ref{fig08} are remarkably close to each other, but although
regions beyond the turn off have not been much affected by the inclusion of the binaries, 
the fact that the main sequence has shifted out of correspondence with these other regions has
resulted in poorer fits, and although no systematic offsets in age, metallicity or reddening correction
have appeared, the error ellipses have certainly increased in size.

The particular case of blue stragglers has not been treated, but a good approximation can be obtained from the
binary contamination example included. Binaries and blue stragglers can populate the region close to the 
turn off, and result in poorer fits. However, having weighted all the CMD equally, the much larger main sequence
population, and the phases beyond the turn off, where errors are much smaller, anchor to a large extent the fit
against a small number of blue stragglers or binaries very close to the turn off. More than 2 local error sigma 
away from the turn off, you can fill in almost any number of blue stragglers, by construction their effect will be 
negligible, see Eq.~\ref{eq:G}.

In the context of inverting resolved stellar populations, a binary does not necessarily imply
a physical gravitational association between two stars, but merely that the light of two 
individual stars has appeared in the data as coming from a single point. The two stars hence show as a single
one having the total luminosity of the sum for the two stars, and an effective temperature
being the energy weighted sum of the temperatures of the two stars, as per a Stefan-Boltzmann law.
Therefore, especially if thinking of dense systems such as globular clusters where crowding effects can be
dominant, the adoption of the same IMF for the secondary component of the binary as for the
primary one is justified.

\subsection{Effects of different sets of isochrones}

We now turn to a possibly important systematic, related to the validity of the underlying 
hypothesis that a single real star from an observed CMD does in fact behave and evolve
as the stellar evolutionary code one is using to construct an isochrone grid assumes it 
to. 
Figure \ref{fig09} gives the CMD produced again from the same point  $(10.9, -3.05, 18.8, 0.05)$
in parameter space, but using this time the Padova stellar evolutionary models (Girardi et al. 2002), 
whilst in all previous examples, the Yale isochrones where used (Demarque et al. 2004).
The distribution of stars on the
magnitude-colour plane is much the same as in Figure \ref{fig03}, and no significant differences can 
be detected.

A number of such CMDs where then inverted using the Yale isochrones, 
resulting in the error ellipsoid whose projections onto the 6 usual planes is shown in
Figure \ref{fig09}. The most immediate conclusion is that changing the stellar models has resulted
in no systematics beyond the  $1\sigma$ level, beyond a slight offset in the recovered reddening
correction of around 0.02 magnitudes. Asides from the above, comparing with Figure \ref{fig03} we note
that using a 'wrong' isochrone grid in the inversion procedure has resulted in slightly
poorer fits, evident in larger error bars on the recovered parameters, the extremes of the
 $1\sigma$ confidence interval now span close to 2.0 Gyr, as opposed to only 1.2 Gyr
when the same isochrone set was used to construct the CMDs and to invert them. This has been due to
the details of the stellar evolutionary code in terms of the differential duration of stellar phases, 
more than to changes in the isochrones as plotted on the CMDs. In this particular case, the error ranges
for the metallicity have grown by a larger fraction than the others. However, this is not a general feature of changing the
isochrone set, but something which depends on the errors, no. of stars considered, actual age, distance and metallicity of
the SSP being studied. Regarding the error ranges, the examples we have shown are only illustrative, and trustworthy error ranges
have to be internally calculated for each particular case.

Note also that these different sets of isochrones use different calibrations to go from the
theoretical diagram to the observed ones in the different sets of filters. This is perhaps
the largest source of uncertainty at this level (VandenBerg 2007) for this range of masses
and ages.

\figce{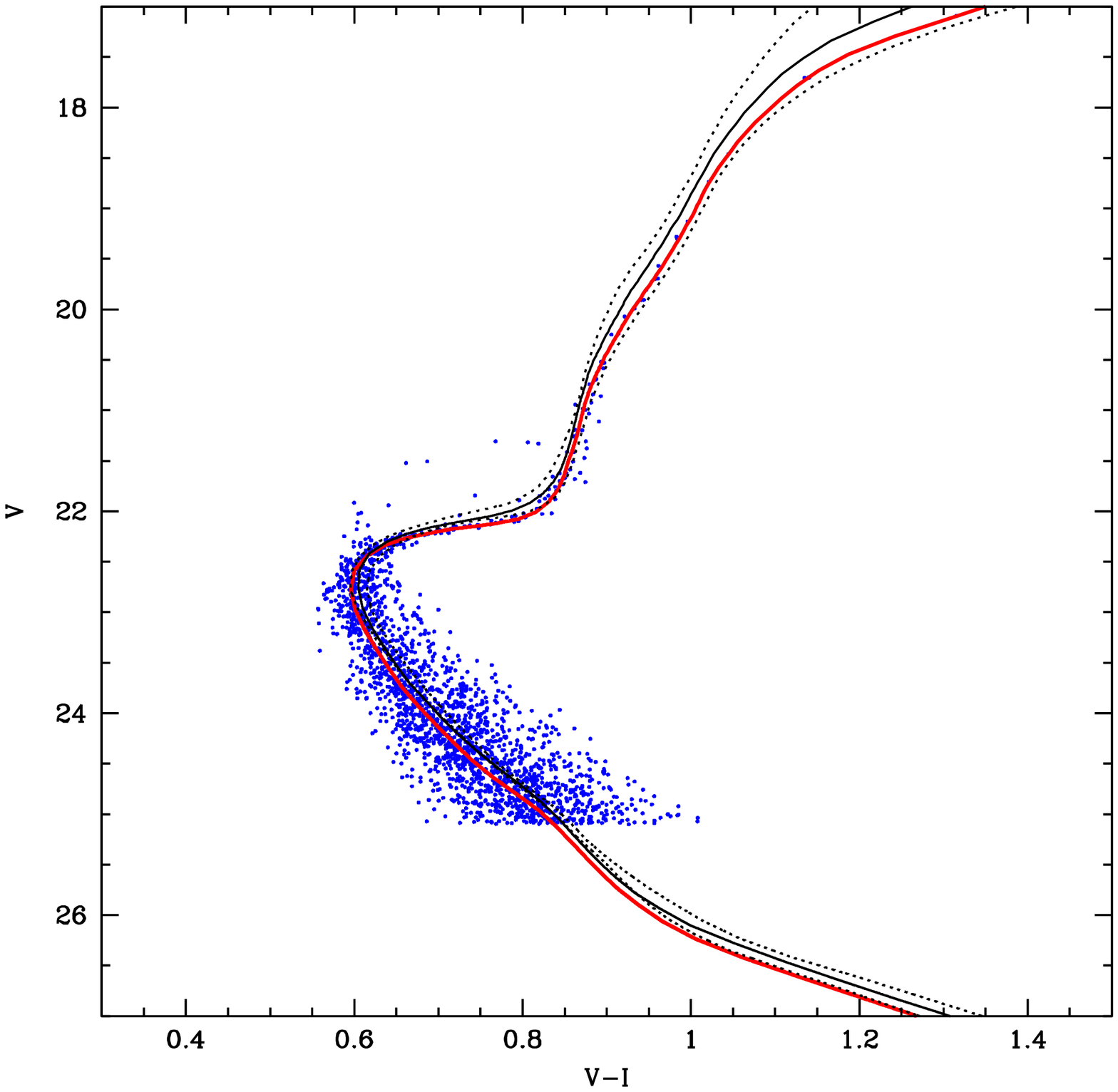}{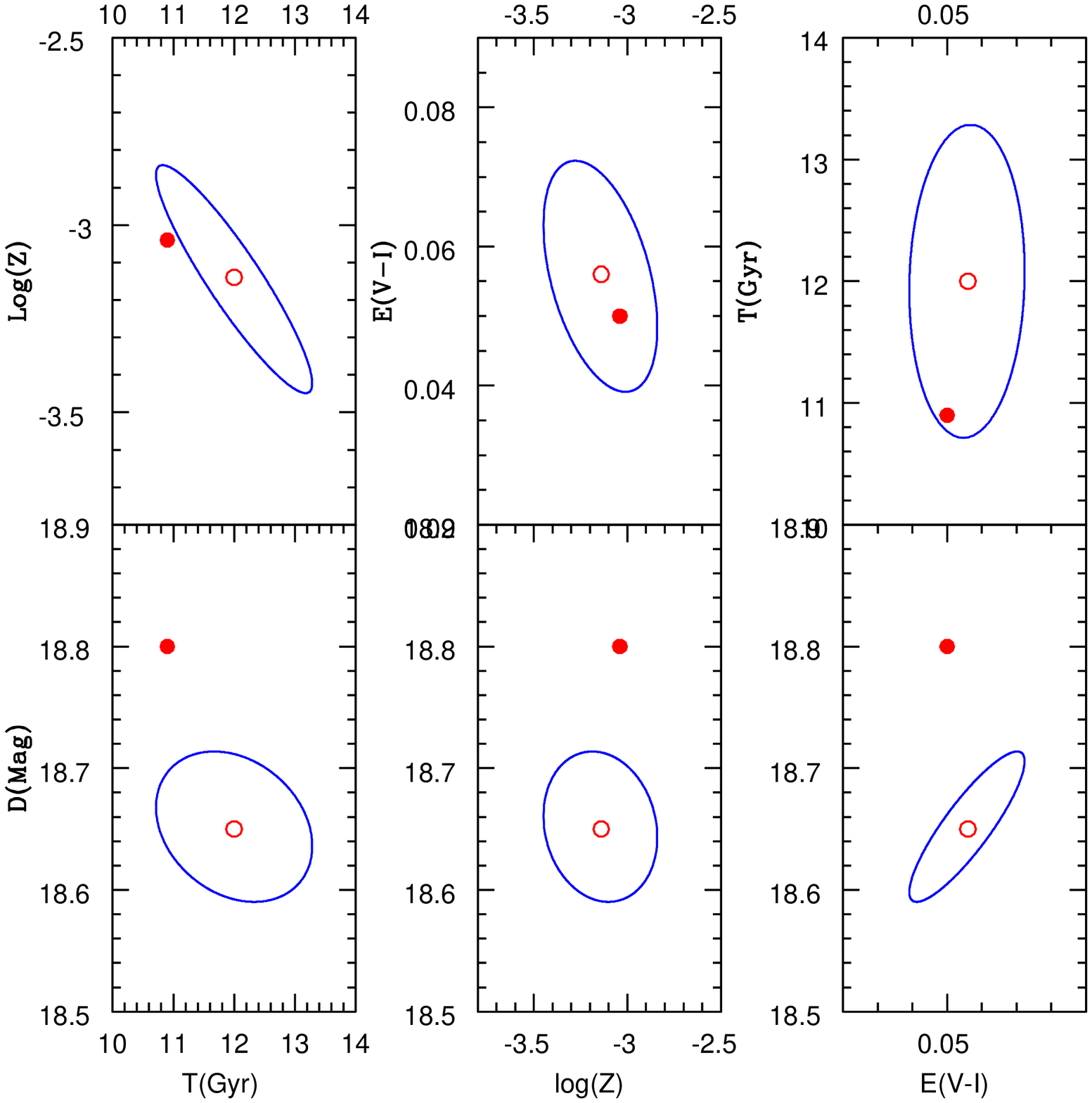}{Simulated Single Stellar Population CMD having an age of 10.9 Gyr and 
$\log Z=-3.05$. The thick (black) curve shows the input isochrone which was used 
adding an unresolved binary companion to 50\% of the stars, with the 
optimal fit isochrone recovered by the method, assuming no binaries, appearing as 
a thin (red) 
curve. Dotted curves show the youngest and oldest isochrones allowed by
the method, at the $1\sigma$ level, which in fact span a 1.52 Gyr interval.}{The six panels show the projection of the error
 ellipsoid resulting from the Monte Carlo
simulation and inversion of a SSP with age 10.9 Gyr and metallicity
$\log Z=-3.05$, onto different planes, for CMDs constructed including 50\% unresolved binaries
and inverted assuming no binaries. The filled circles show input parameters, and
the empty ones the average values for the recovered parameters.}{fig08}

All the isochrones throughout have been used only up to the tip of the RGB, with regions beyond which give rise 
to interesting 'red clump' morphologies, having been excluded from the analysis. 
Even though inclusion of these phases, for example the HB region, would provide interesting
independent restrictions and potentially enhance the effectiveness of the method, at present
post helium flash regions have to be modeled through the inclusion of RGB mass loss rate 
recipes which can not yet be calculated or constrained from first principles. Thus,
their inclusion would necessarily imply obtaining inferences which would be limited to the
validity of the currently poorly grounded hypotheses one would chose regarding late stellar
evolution.

Perhaps we can not conclude
that stellar evolutionary modeling has converged to a final solution, even below the helium flash. 
However, by using broad band colours,
where integration over relatively wide sections of a stellar spectrum is included, an averaging over 
details in fine features, and by restricting the analysis to stellar phases which are well understood,
a method such as what we have implemented here is now capable of yielding results which are highly robust with
regards to the current discussions and disagreements over stellar evolution.

Certainly in going from Figure \ref{fig03} to Figure 9 the error ranges
have increased, but the inferences correctly identify ages, metallicities and distances to within
internally computed confidence intervals. In future, it would be desirable to use stellar evolutionary
models more extensively, including phases beyond the helium flash, to yield much more restrictive
inferences and narrow error ranges, but whilst debate rages over the physics of stars in those
regions, we prefer to increase the size of both our error ranges and our confidence in the final results.
The same applies to the use of more detailed features of stellar spectra, in going from broad
band colours to narrow band colours and finally details of the emission line structure, more
information is encoded, the potential of more refined inferences is there, but at least until the 
different models converge, inferences using such details should be considered as subject to
short term revision.

\subsection{Sampling and background/foreground contamination}

As remarked previously, the numbers of stars in an observed CMD and the size of the observational errors 
dispersing the stars from their intrinsic locations all influence the quality of the inferences
which can be obtained. Figure \ref{fig10}  is analogous to  Figure \ref{fig03}, the 
same point in parameter space has been modeled, with the same errors, but the number of stars has
been reduced in half, of the $\sim 2000$ stars simulated in all previous examples, only $\sim 1000$
where simulated this time.

%%% fig 09

\figce{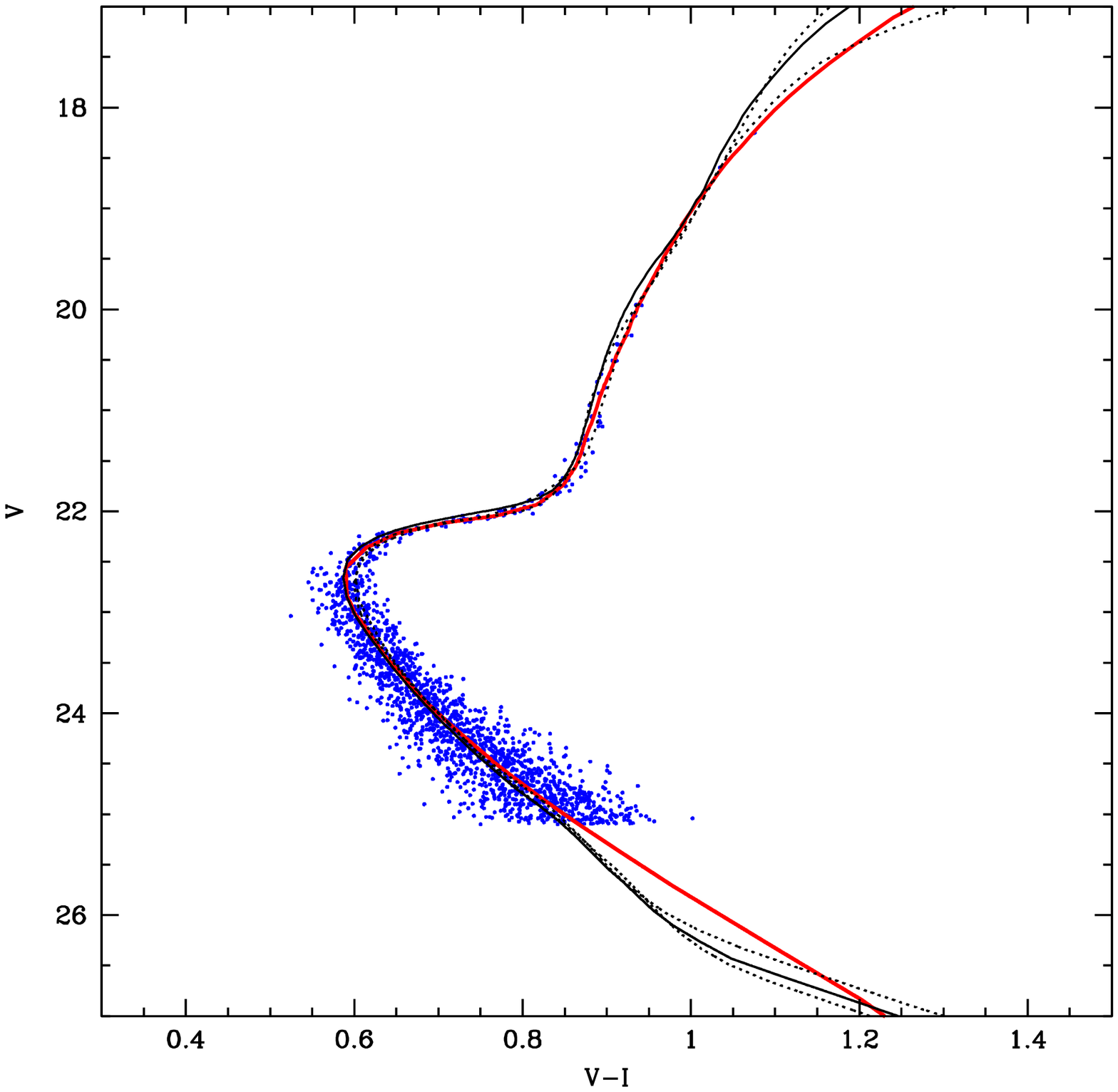}{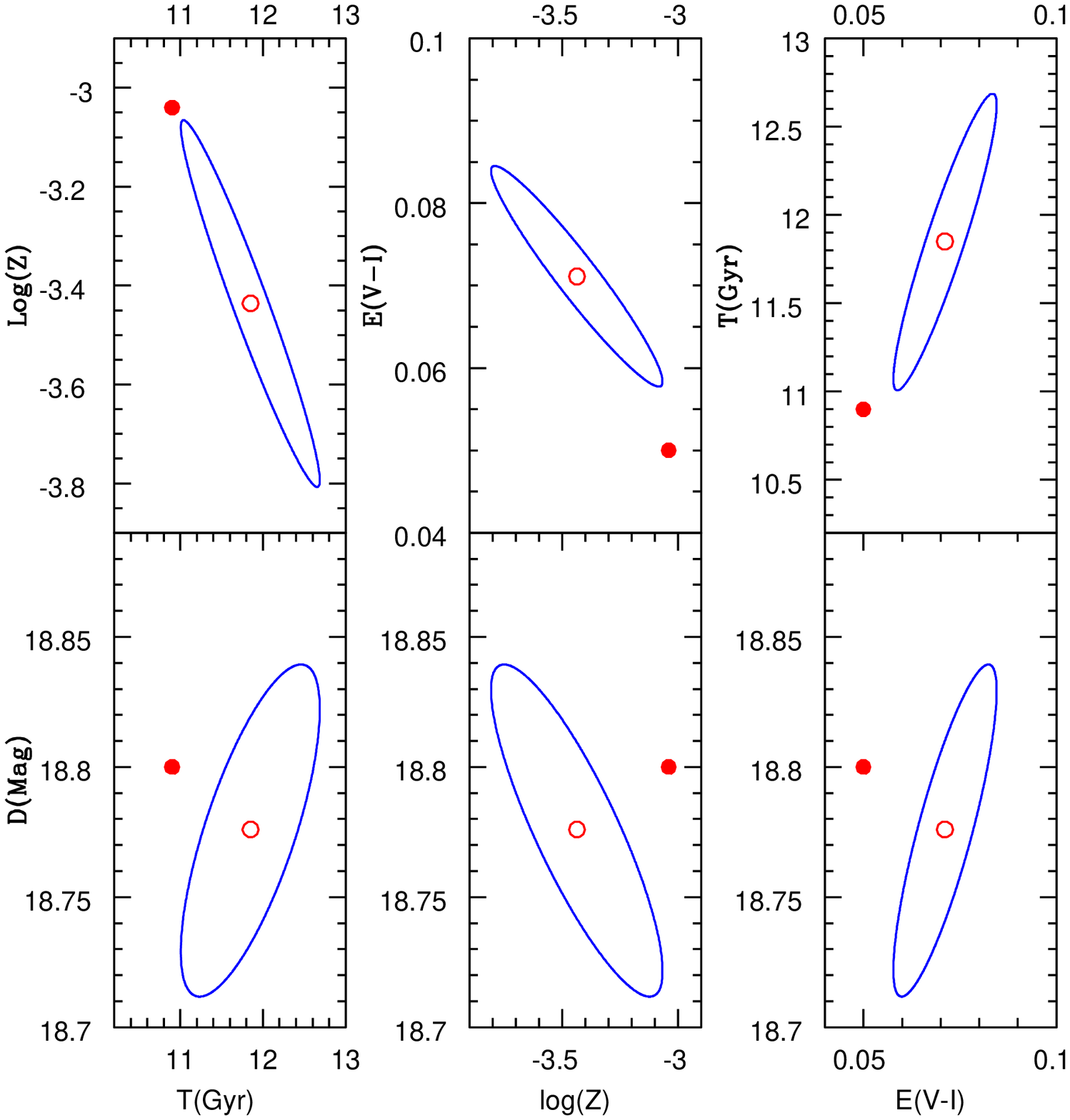}{
Simulated Single Stellar Population CMD having an age of 10.9 Gyr and 
$\log Z=-3.05$. The thick (black) curve shows the input isochrone (Padova models), with the 
optimal fit isochrone (Yale models) recovered by the method, appearing as a thin (red) 
curve. Dotted curves show the youngest and oldest isochrones allowed by
the method, at a $1\sigma$ level, which in fact span a 1.8 Gyr interval.}{The 
six panels show the projection of the error ellipsoid resulting from the Monte Carlo
simulation and inversion of a SSP with age 10.9 Gyr and metallicity
$\log Z=-3.05$, onto different planes, for CMDs constructed using Padova models
and inverted using Yale ones. The filled circles show input parameters, and
the empty ones the average values for the recovered parameters.}{fig09}

We see from Figure \ref{fig10} that no systematics appear, and that even with only $\sim 1000$
stars, a number easily achievable in current observations of Galactic Single Stellar Populations,
the recovered parameters are all within the $1\sigma$ error ellipses of the input ones. 
By comparing with Figure \ref{fig03} we do note
that again, as in the two previous experiments, the confidence intervals have grown considerably,
a similar trend is seen if the assumed observational errors are increased. The CMD of Figure \ref{fig10} 
is clearly much more thinly populated than previous ones, but still the method manages to
recover isochrones that do not differ much in their parameters from the input ones.

\begin{table*}
 \centering
 \begin{minipage}{150mm}
  \caption{Summary of the systematics effects explored. The recovered (output) values include 
their corresponding 1$\sigma$ confidence intervals.}
  \begin{tabular}{lcclclll}
\hline
 Test     &   Figure    &   $t$(input) (Gyr)  &   $t$(output) (Gyr)  &   $\log Z$ (input) 
 &   $\log Z$ (output)    & \multicolumn{1}{c}{   $D$ (mag)$^a$}    &
\multicolumn{1}{c}{      $R$ (mag)$^b$ }     \\
\hline
 Old metal rich SSP   &  1  & 13.90 &13.95 $\pm$ 0.50 & $-2.70$ &$-2.95 \pm 0.38$ & 18.845 $\pm$ 0.06 & 0.082 $\pm$ 0.015 \\
 Typical old SSP      &  3  & 10.90 &11.65 $\pm$ 0.74 & $-3.05$ &$-3.21 \pm 0.17$ & 18.795 $\pm$ 0.04 & 0.059 $\pm$ 0.015 \\
 Old metal poor SSP   & 4  & 13.90 &13.20 $\pm$ 0.72 & $-4.00$& $-3.80 \pm 0.21$ & 18.775 $\pm$ 0.04 & 0.051 $\pm$ 0.008 \\
 Young metal rich SSP &  5  & 6.90 &6.950 $\pm$ 0.50 & $-2.05$& $-2.03 \pm 0.06$ & 18.780 $\pm$ 0.06 & 0.045 $\pm$ 0.012 \\
 Young metal poor SSP &  6  & 6.90 &7.100 $\pm$ 0.50 & $-3.37$& $-3.41 \pm 0.13$ & 18.785 $\pm$ 0.03 & 0.050 $\pm$ 0.006 \\
 Wrong IMF           &  7  & 10.90 &11.25 $\pm$ 0.45 & $-3.05$ &$-3.13 \pm 0.20$ & 18.795 $\pm$ 0.06 & 0.051 $\pm$ 0.012 \\
 50\% binaries      &  8  & 10.90 &12.00 $\pm$ 1.26 & $-3.05$& $-3.15 \pm 0.30$ & 18.650 $\pm$ 0.06 & 0.056 $\pm$ 0.015 \\
 Wrong isochrones    &  9  & 10.90 &11.90 $\pm$ 0.90 & $-3.05$& $-3.44 \pm 0.38$ & 18.775 $\pm$ 0.06 & 0.070 $\pm$ 0.014 \\
 50\% less stars     &  10 & 10.90 &11.40 $\pm$ 1.05 & $-3.05$& $-3.13 \pm 0.25$ & 18.772 $\pm$ 0.07 & 0.049 $\pm$ 0.021 \\
 60\% contamination  &  11 & 10.90 &12.08 $\pm$ 1.39 & $-3.05$& $-3.12 \pm 0.41$ & 18.751 $\pm$ 0.12 & 0.061 $\pm$ 0.023 \\ 
\hline
\end{tabular}
\noindent $^a$ The input distance modulus was kept at 18.80 mag in all cases.\\
\noindent $^b$ The input reddening was kept at 0.05 mag in all tests.
\end{minipage}
\end{table*}

\figce{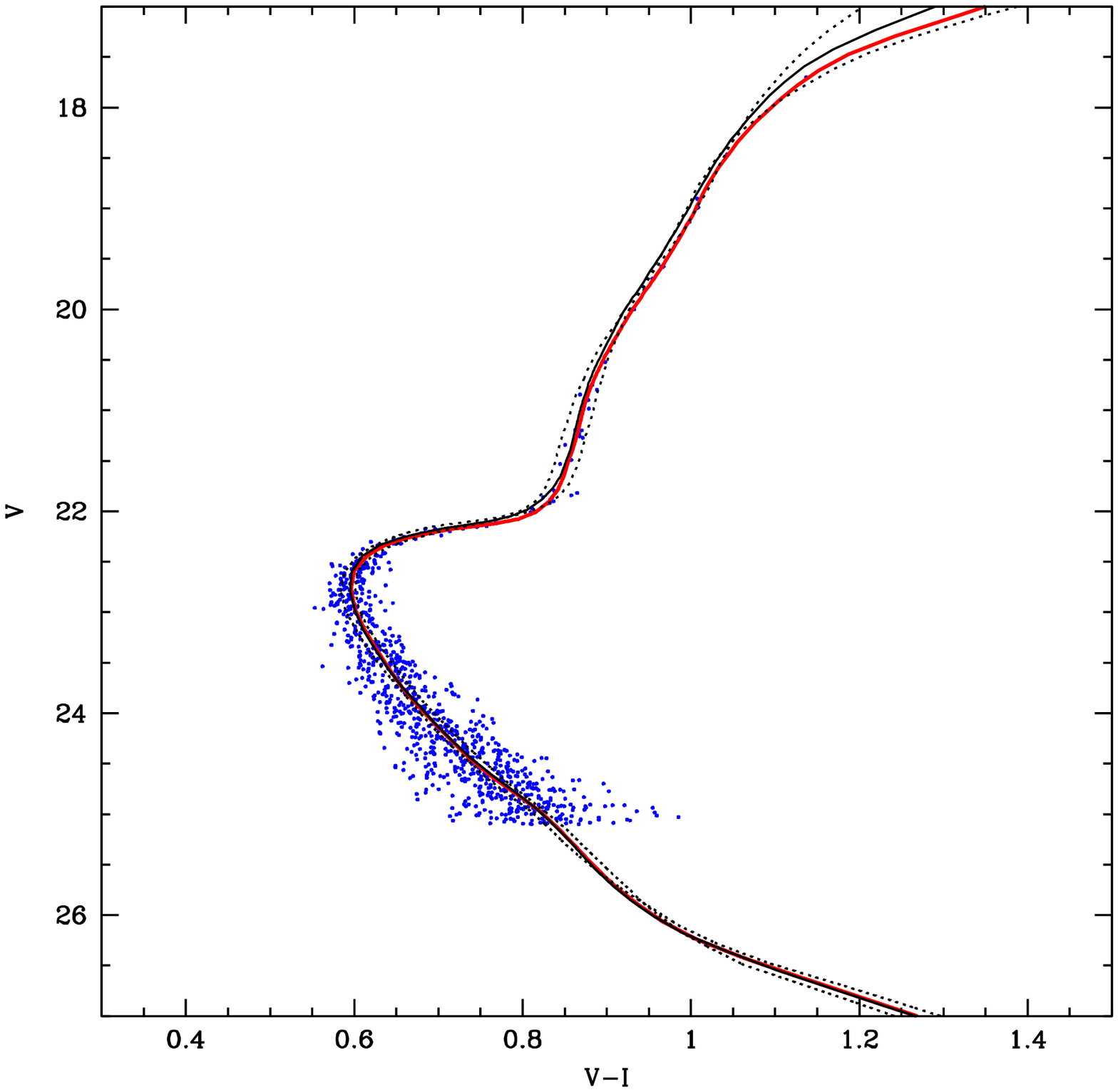}{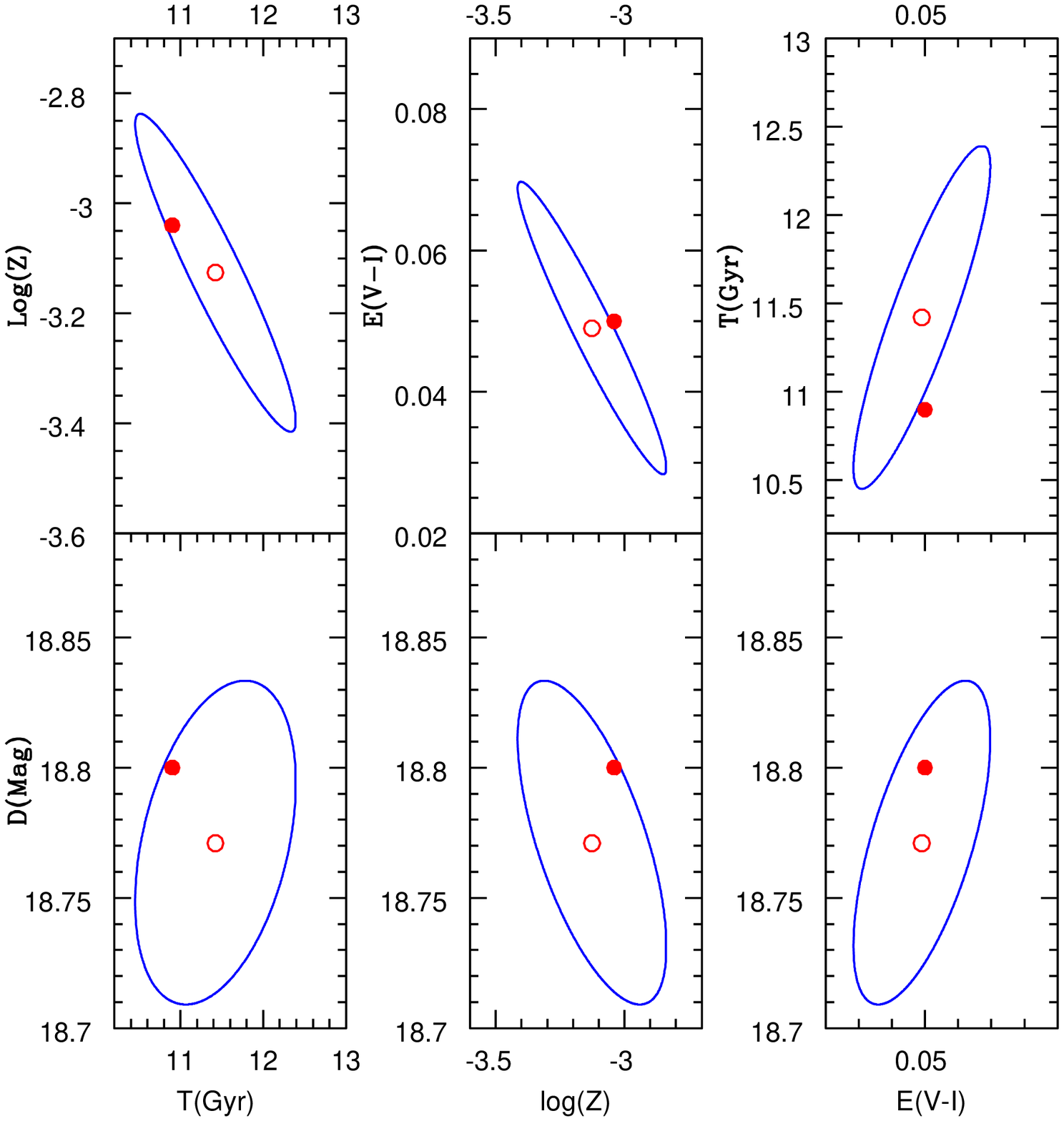}{Simulated Single Stellar Population CMD having an age of 10.9 Gyr and 
$\log Z=-3.05$. The thick (black) curve shows the input isochrone, with the 
optimal fit isochrone recovered by the method, appearing as a thin (red) 
curve. Dotted curves show the youngest and oldest isochrones allowed by
the method, at the $1\sigma$ level, which in fact span a 2.1 Gyr interval.}{The six panels show the projection of the error 
ellipsoid resulting from the Monte Carlo
simulation and inversion of a SSP with age 10.9 Gyr and metallicity
$\log Z=-3.05$, onto different planes, for CMDs constructed using only $\sim 1000$ star as opposed to 
the $\sim 2000$ used in all previous examples. The filled circles show input parameters, and
the empty ones the average values for the recovered parameters.}{fig10}

As a final test of systematics which surely apply to the use of real data, we 
present Figure \ref{fig11}, where a fraction of 60\% (by number) of polluting stars have been added to the CMDs,
also produced from the same parameters as the previous tests. The presence of polluting stars is 
evident in Figure \ref{fig11} where the area in the CMD which is taken into account in the analysis is
clearly seen to have been filled with contaminating stars. To make this example as realistic as possible, 
we have modeled the contamination produced by field stars using the Besan\c{c}on model of the Galaxy 
(Robin et al. 2003). We simulated a field of one square degree along the 
direction $(l,b) = (45\deg, 23\deg)$ 
so that a significant contribution by the thin and thick discs is present. The errors which we have assumed
apply to the polluting points are the same as the average for the actual 'observed' stars at a given 
luminosity. From the 87000 simulated stars along that direction, we extracted a random sample of size 1200 
to simulate a severe contamination of 60 \% with respect to the number of stars in the CMD of the simulated cluster. 
This contamination appears in the CMD as an essentially constant density addition of random stars, as the
typical scales over which variations appear in the Besan\c{c}on model in both colour and magnitude, 
are larger than the domains of Figure~\ref{fig11}.

From the integral over the model isochrone being compared to the data shown in 
equation \ref{eq:Gint}, and from the form of $G_{i}(T,Z,D,R)$ of equation \ref{eq:G}, it is obvious that
a given star will only contribute significantly to ${\cal L}(T,Z,D,R)$ for isochrones
passing within 1-2  $( \sigma(L_{i})^{2} + \sigma(C_{i})^{2})^{1/2} $ of it. The convolution
of the error Gaussian which maps the model stars onto the CMD and the error Gaussian for an
observed star converges quickly within a distance of  $2\sigma$ of either. This means that
features not yet well modeled by the theoretical stellar models, but which fall directly on the isochrones,
such as the Red Giant Branch bump morphology seen in luminosity functions of RGBs, will not have any effect at all on our inferences.
The CMDs we have modelled do not have sufficient stars to properly allow us to identify any Red Giant Branch bump,
however, this feature is a second order effect the paucity of which, together with the fact that it does not move stars out of
the isochrone, and only slightly changes the details of the density of points along it, that make the method
robust against it. The uncertainties in our inferences due to the level of observational errors are already
much larger than anything theoretical Red Giant Branch bump uncertainties might introduce.
Similarly, the presence of a small number of blue stragglers will tend to bias the
estimation, but given that their number is small in comparison with all the other
stars which lie along the tentative isochrone, they do not bias the final estimate.

The above discussion leads us to expect that the method we have constructed should be highly
robust against the presence of contaminating stars, as indeed can be concluded from Figure \ref{fig11}, where the resulting
error ellipses for this experiment are shown. We can see that although the errors have increased,
all inferred parameters are practically within $1\sigma$ error ellipses of the input ones. 
Some inferences are more affected than others, for example, the error in the age determination has
nearly doubled when compared to Figure \ref{fig03}, but the one in the reddening correction
has hardly changed. This result holds even for large polluting fractions than what is being simulated, and in fact suggests
that the method can be applied to the identification of stellar populations in fields crowded by
foreground and background contamination. The method is however sensitive to 'polluting' stars occurring
systematically close to critical regions of the isochrones, e.g. a sequence slightly off to one
side of the RGB corresponding to the observed main sequence, essentially equivalent to having used
grossly mistaken stellar tracks. The precise effect of foreground/background contamination will in fact depend
on the details of the SSP being studied, $(T, Z, D, R)$ parameters, numbers of stars and observational errors,
and location of contaminating stars. The final Monte Carlo assessment of the confidence intervals must hence also 
include a modeling of the fields contamination, in order to yield meaningful results.

Lastly, as a summary of these last 2 sections we present Table 1, where all the synthetic cases we have treated are given. The
values for $(T, Z, D, R)$ in the table give the input parameters, followed by the recovered ones with the corresponding
1 sigma error ranges.

\section{An example with observations}

In this final section we present an example of the workings of the method, using data
from an observed SSP, where in a sense, all possible systematics are at work.
Of the 7263 stars observed by Rosenberg et al. (2000) for the globular cluster NGC~3201, we take the 5197 stars brighter
than $V=20$, together with the corresponding error estimates, and use them as inputs for the method.
The value of the magnitude cut used for this example was chosen so that the example should be comparable to
the synthetic cases treated previously, where a lower cut of 2 magnitudes below the turn off was consistently
assumed. This value guarantees the level of accuracy seen in the synthetic examples, and is designed to exclude the lower main
sequence, where incompleteness begins to enter the data. In this particular case, the completeness limit
given by the observations lies below the lower limit we take.  Also, the discarded region is one of
inherent low information content (see discussion following Equation~\ref{eq2}). 
By construction, the method assumes the observations are complete down to the lower limit
included. If one is forced to cut the CMD closer to the turn off region, the quality of the fit is reduced, with
growing error ellipses. Inclusion of regions beyond the completeness limit will result in formal error ranges
which can appear smaller, but violation of the hypothesis of completeness 
will result in unreliable inferences.

\figce{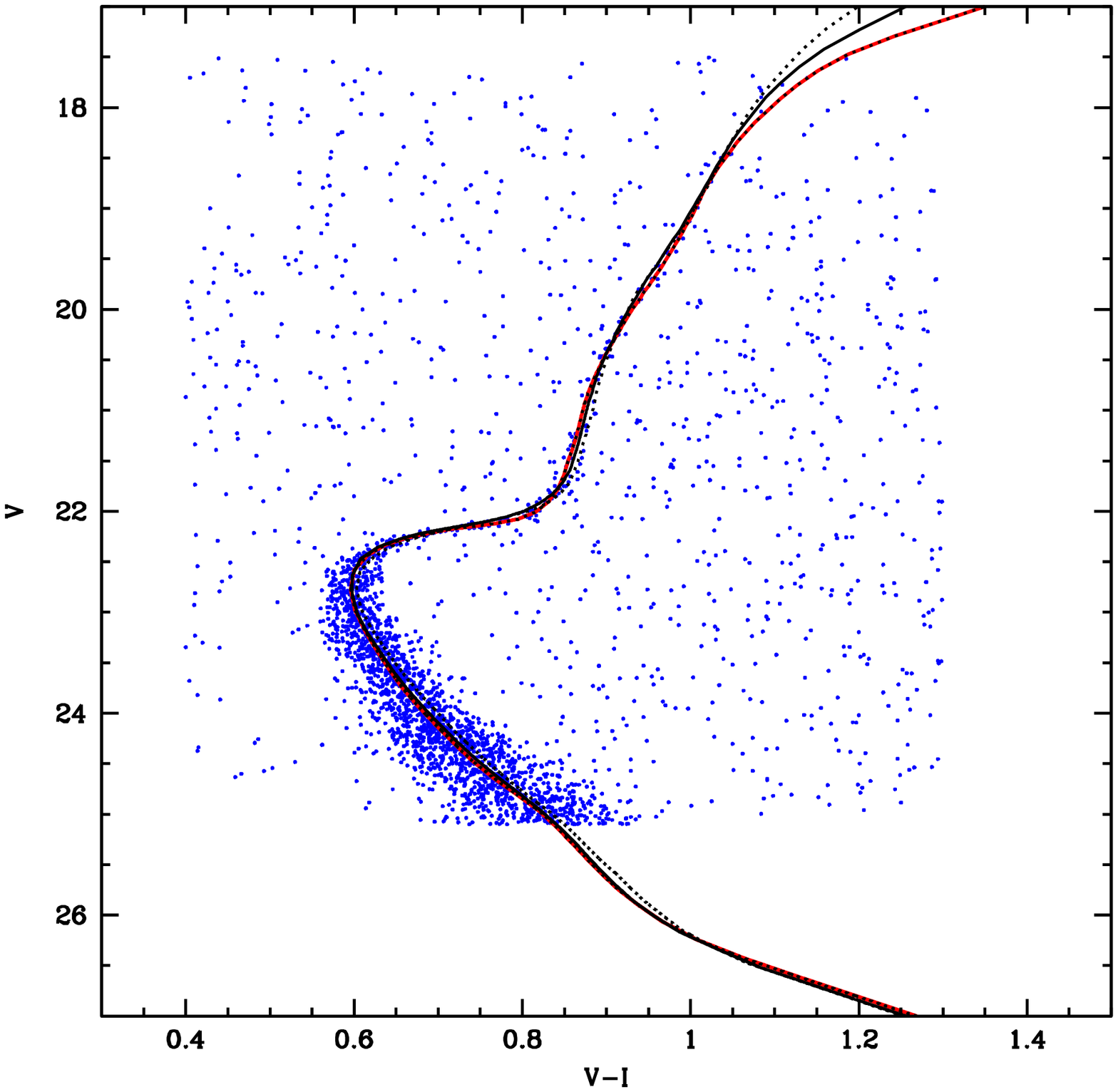}{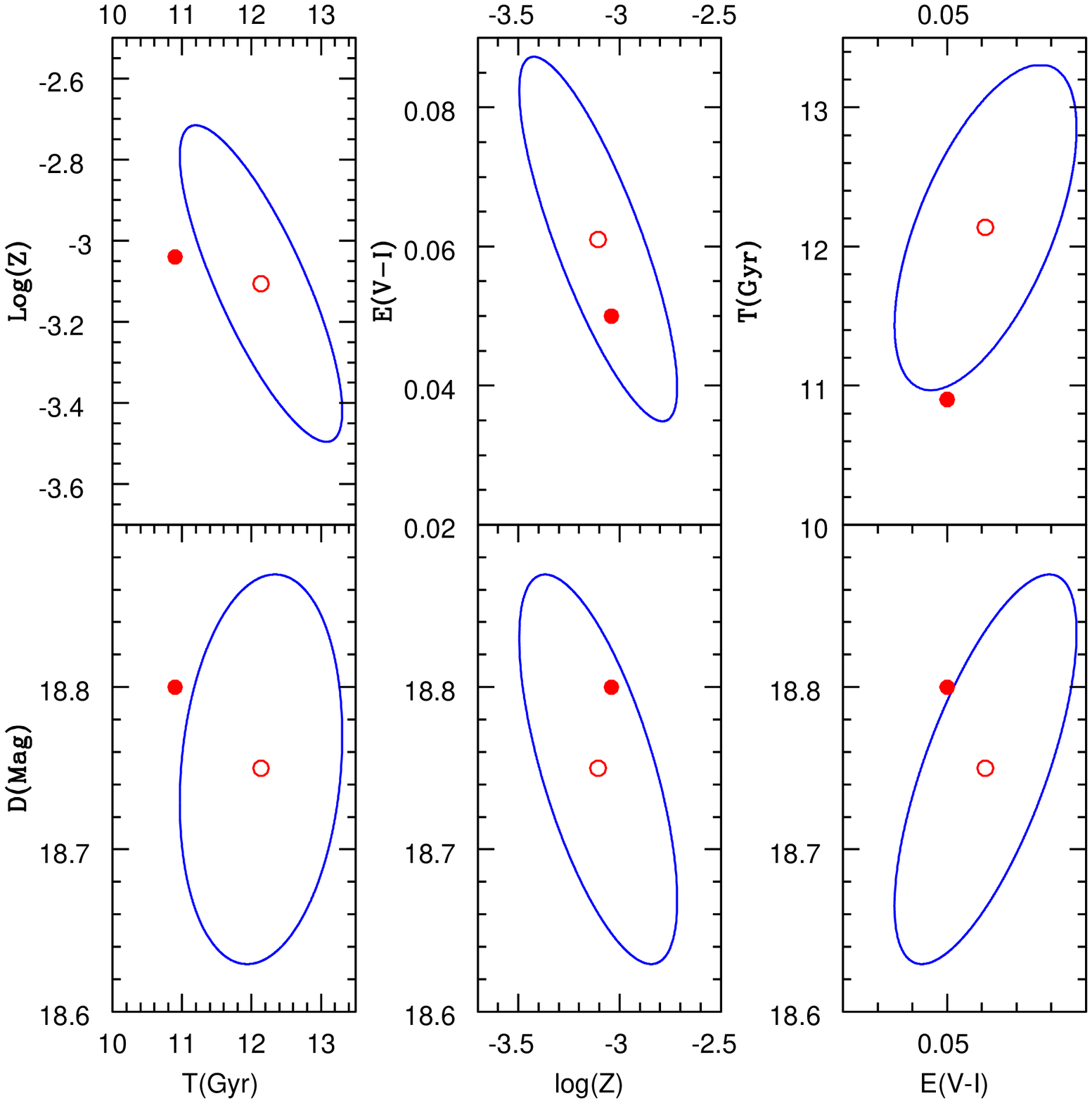}{Simulated Single Stellar Population CMD having an age of 10.9 Gyr and  
$\log Z=-3.05$. The thick (black) curve shows the input isochrone, with the 
optimal fit isochrone recovered by the method, appearing as a thin (red) 
curve. Dotted curves show the youngest and oldest isochrones allowed by
the method, at the $1\sigma$ level, which in fact span a 1.78 Gyr interval.}{The six panels show the projection of the error 
ellipsoid resulting from the Monte Carlo
simulation and inversion of a SSP with age 10.9 Gyr and metallicity
$\log Z=-3.05$, onto different planes, for CMDs including the addition of 60\% Galactic
pollution stars. The filled circles show the input parameters, and
the empty ones the average values for the recovered parameters.}{fig11}

In general, the choice in the details of any implementation will depend on the robustness of the available data
and theoretical models. One would expect that with time, stellar modeling will improve to the point
of making more reliable predictions, not only for very young stellar populations, but also for stellar phases
beyond the helium flash, perhaps even into the white dwarf cooling sequence, and including all the currently
poorly understood features like blue stragglers and horizontal branch. At that point, the implementation can be
trivially updated by simply changing the range of masses considered along the theoretical isochrone, yielding
more accurate and restrictive inferences. On the observational side, when for example GAIA data are available,
the effectiveness of the implementation can be improved by including stars further down along the main sequence,
to result in improved metallicity and reddening corrections. 

The actual stars for this cluster appear in the left panel of Figure \ref{fig12}, together with the 
best fit isochrone found by the method, using the Yale isochrones, changing the isochrone set, yields
results only marginally different. This is clearly seen to be a good fit, 
obtained this time through a completely automatic and objective procedure, maximizing use of information 
both from the data and from the stellar models with which the data was compared.

The basic parameters associated to this best fitting isochrone,
together with their 1-$\sigma$ confidence intervals are: $T=12.1 \pm 0.5$ Gyr, $\log Z=-3.14 \pm 0.15$, 
$D = 14.03 \pm 0.1$ mag,  and E(V-I)$= 0.34 \pm 0.02$ mag, and are shown in the right panel of 
Figure \ref{fig12}. As in the previous 
examples, the method returns not only best fit values and associated statistically robust confidence
 intervals,
but also the full covariance structure of the inference, as shown in Figure \ref{fig12}. We see the 
1-$\sigma$ confidence 
ellipse in the $(T,Z)$ plane slanting as expected from the 'age-metallicity degeneracy', but again, almost
no correlation in the $(T,D)$ plane. Again, when moving along the correlations in the recovered 
parameters, we see
that the maximum and minimum age isochrones, at the 1-$\sigma$ level, dotted curves, are practically 
indistinguishable from 
each other, or from the the best fit one, based only on shape. The three solid dots in 
Figure \ref{fig12}
 give the center of our inferences, for three different values of the [$\alpha$/Fe]
 enhancement parameter, 0.0, +0.3 and +0.6,
a final systematic, which also has little effect\footnote{We used the approximation given by 
Pagel (1996) to transform the observed [Fe/H] abundance into a metallicity.}.
 The isochrones shown are only for the alpha enhancement
of 0.0, as the inclusion of the corresponding isochrones for the other two values
 would only result in a complete
overlapping of curves, all within the region defined by the oldest and youngest 
isochrones at [$\alpha$/Fe] =0.0  
shown. The resolution of the method, as mentioned earlier, 
is much enhanced from a simple geometric fitting of isochrones by having included the duration of 
evolutionary 
phases explicitly in the merit function being used. Confidence intervals similar to the case 
of Figure \ref{fig11} are the 
result of the large number of stars available above the lower magnitude cut.

\figce{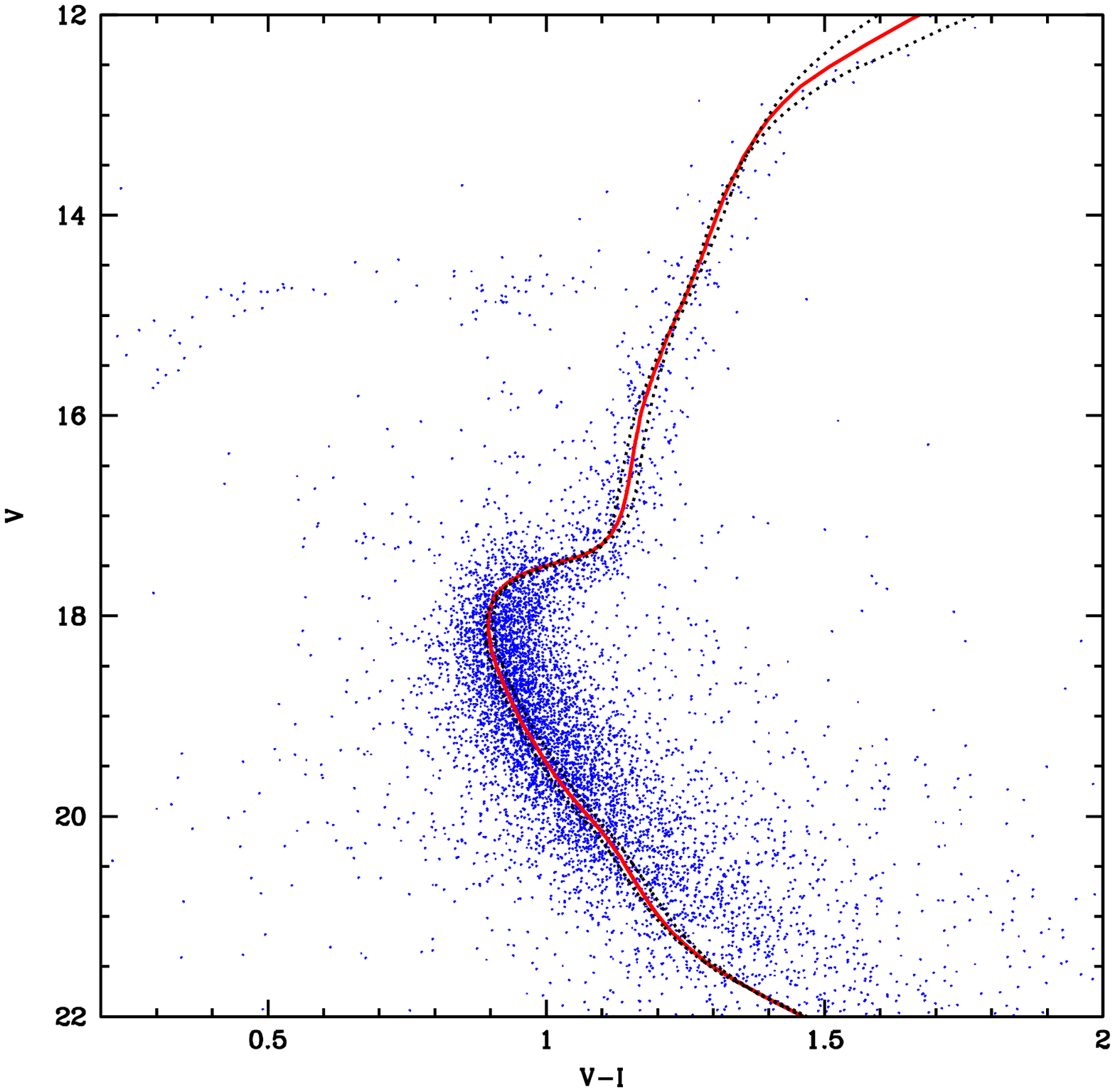}{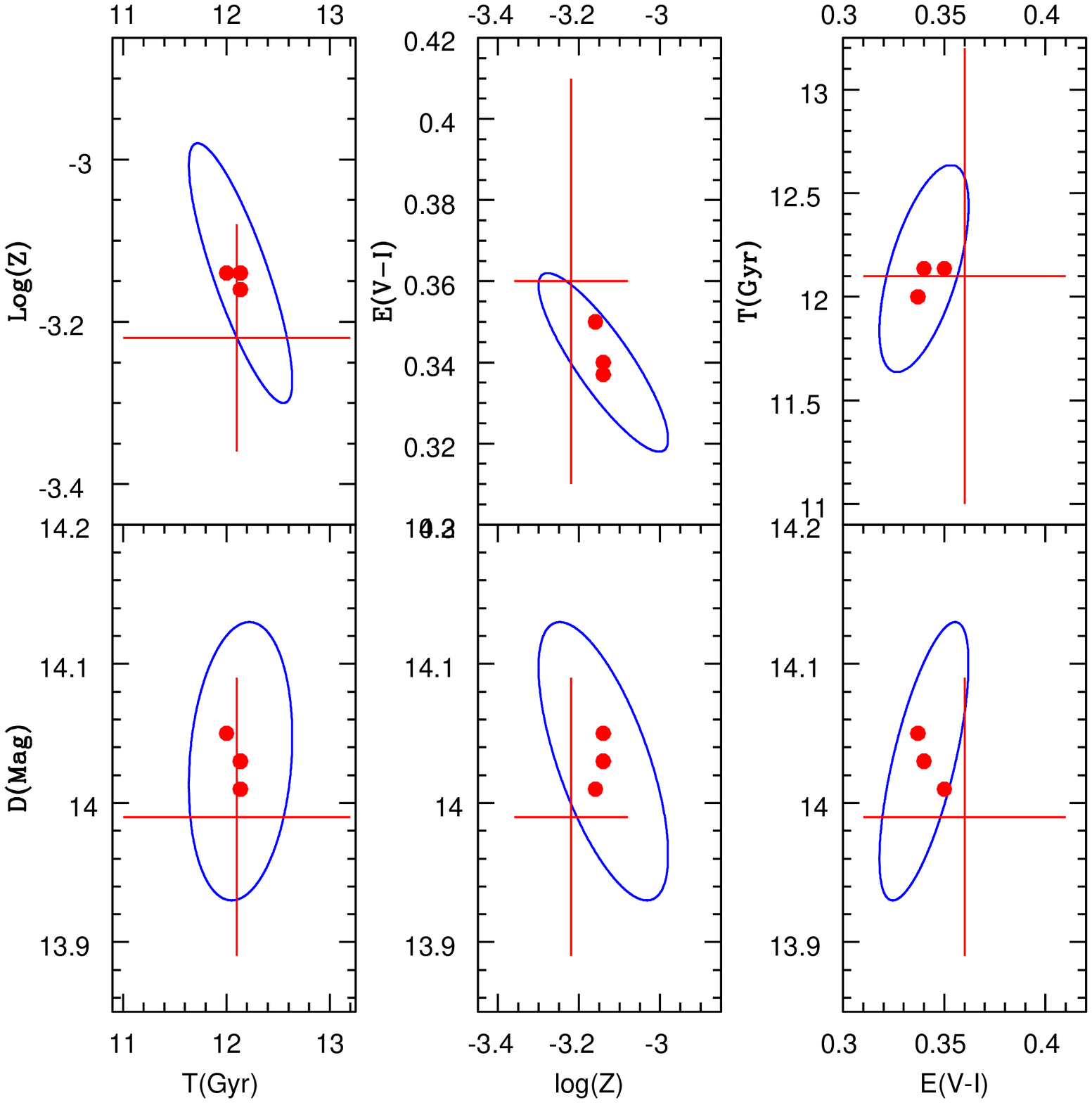}{Observed CMD for NGC~3201. The thick (red) curve shows 
the optimal fit isochrone 
recovered by the method. Dotted curves show the youngest and oldest isochrones allowed by
the method, at the $1\sigma$ level, which in fact span a $1.0$ Gyr interval.}
{The six panels show the projection of the error 
ellipsoid resulting from the Monte Carlo
simulation and inversion of a SSP with age 12.1 Gyr and metalicity
$\log Z=-3.14$, onto different planes, for CMDs including a similar number of stars, and errors
modeled as the reported ones for NGC~3201. Filled circles show the average values for the 
recovered parameters, for 3 different values of assumed [$\alpha$/Fe] enhancement, 0.0, +0.3 and +0.6, and the
crosses give inferences and uncertainties as found in the recent literature for this cluster, see text.}{fig12}

For comparison, in the  globular cluster compilation by Harris (1996), NGC~3201 is listed as having
$(m-M)_{V}=14.24$ mag, E(B-V) = $0.24$ mag (which translates into E(V-I)=0.25), and a metallicity of
$\log Z=-3.28$, with no confidence intervals for any of the above.
The more recent Salaris \& Weiss (2002) report an age for this cluster of
$12.1 \pm 1.1$ Gyr, Kraft \& Ivans (2003) quote a metalicity of between $\log Z=-3.08$ and $-3.36$, 
taking the extremes
of all the possibilities they give, for different red giant spectral metalicity determinations and a solar value of 
$\log Z_{\odot}=-1.7$. Rosenberg
et al. (2000) give a value for the apparent $D=(m-M)_{V}=14.17$ mag, with no estimates of the 
associated confidence 
intervals. Recent determinations of the reddening towards this field include Salaris \& Cassisi (1998) 
giving
E(B-V) = $0.24 \pm 0.02$ mag (which translates into E(V-I)=0.29 $\pm 0.024 $), and 
Piersimoni et al. (2002) giving E(V-I)=$0.36 \pm 0.05$ mag, which compares very well with
our measure of E(V-I)=R=$0.34\pm0.02$. This results in a distance modulus of $\mu = D - 1.92$E(V-I) = 
13.38$\pm$0.1, which agrees with the determination of Piersimoni et al. (2002) of $\mu = 13.31\pm0.06$ mag
based on RR Lyr stars in the cluster. 

The crosses in Figure \ref{fig12} summarize the independent determination for the parameters of NGC~3201
which we treated, as discussed in the text above. Overall, comparing our results for NGC~3201
with independent determinations from the literature, we find very good agreement within the uncertainties 
of the different determinations, and we notice that our results have typically more restrictive
confidence intervals. We see that in spite of 
 the absence of evolutionary phases beyond the helium flash in the models we used, the fit is robust and 
yields good
results. The presence of a level of contamination and a horizontal branch, lead
to larger error ellipses than in the controlled experiments of previous sections, but no evidently 
wrong systematics
are apparent, we conclude no large systematics remain. The lower region of the CMD appears more poorly 
fitted, a consequence
of having only considered stars brighter than $V=20$ in the fitting procedure. These lower regions of 
the CMD however,
are heavily affected by incompleteness and very large observational errors, requiring, in the context
 of the method
used, their exclusion.

This results can be seen as an overall validation of the method, as metalicities, distances
and reddening corrections reported by the authors cited where often derived through methods completely independent to
the stellar evolution model comparisons we have performed here.
We find that in going to real data, the expectations
of the previous sections hold, and we are able to obtain a best fit solution which is clearly a good solution, 
the resulting isochrone is a good match to the observed stars, and the values for the $(T, Z, D, R)$ vector
obtained match well with completely independent determinations found in the literature.
The fit is of course not perfect, and inconsistencies between stellar models and
reality undoubtedly remain, even below the turn off region, where available stellar models largely agree
with each other. However, the full agreement between our inferences and those derived through
completely independent physical and astronomical methods noted above, show that no systematics are being
introduced by the remaining inadequacies of stellar evolutionary codes, at least not beyond the
confidence intervals internally calculated by the method.

The comparison of results obtained with our method
against accurate independent determinations of metalicities through spectroscopy, or distances through future parallax
measurements, could hence be used to check stellar evolution theory.

\section{Conclusions}

We have presented a method which treats the inference of structural parameters of single stellar populations 
in a fully rigorous statistical manner, within a Bayesian approach, to construct a merit function
for the comparison of an observed CMD in relation to a point in the 4-D parameter space of age, metallicity,
distance modulus and reddening. 

Implementing a genetic algorithm simulation we have shown through various
examples using synthetic CMDs with parameters as appropriate for current observations of
Galactic Single Stellar Populations, that the method accurately recovers the input parameters, and yields
reliable confidence intervals, including naturally an analysis of the
covariances and correlations between the recovered parameters.

By performing extensive Monte Carlo experiments with systematic effects, which surely 
apply to real cases and which are explicitly excluded from the formalism presented, 
we have evaluated the consequences  these effects are expected to have on the results 
of our method. We found that although the error ranges
on the recovered parameters typically increase, no systematic offsets appear, at 
least within the reasonable ranges tested.

This implies that our formalism, which explicitly takes into account the
number density of stars along isochrones, can tentatively test stellar
evolution models when using clusters with well-measured abundances, distances
and reddenings. This, along with the absolute dating of clusters with precise
photometry, will be presented in the second paper of this series (Valls--Gabaud 
\& Hernandez 2007).

\section*{Acknowledgments}
The authors wish to thank the Padova, Yale-Yonsei and Victoria stellar evolution groups 
for assistance and discussions on 
the use of their isochrones, and sharing  material in advance of
publication. The authors wish to thank a second anonymous referee for helpful
comments which improved the final version of the manuscript.
X.H. acknowledges the support of grants UNAM DGAPA (IN117803-3),
CONACyT (42809/A-1) and CONACyT (42748) during the development of this work.
This work was supported by a joint CNRS-CONACYT grant.

\label{lastpage}

\end{document}